\documentclass[fleqn,usenatbib]{mnras}

\usepackage{newtxtext,newtxmath}
\usepackage[T1]{fontenc}
\usepackage{hyperref}
\usepackage[dvipsnames]{xcolor}
\hypersetup{colorlinks,linkcolor={PineGreen},citecolor={PineGreen},urlcolor={PineGreen}} 

\DeclareRobustCommand{\VAN}[3]{#2}
\let\VANthebibliography\thebibliography
\def\thebibliography{\DeclareRobustCommand{\VAN}[3]{##3}\VANthebibliography}

\usepackage{graphicx}	% Including figure files
\usepackage{amsmath}	% Advanced maths commands
\newmuskip\pFqmuskip

\newcommand*\pFq[6][8]{%
  \begingroup % only local assignments
  \pFqmuskip=#1mu\relax
  \mathcode`\,=\string"8000
  \begingroup\lccode`\~=`\,
  \lowercase{\endgroup\let~}\pFqcomma
  {}_{#2}F_{#3}{\left[\genfrac..{0pt}{}{#4}{#5};#6\right]}%
  \endgroup
}
\newcommand{\pFqcomma}{\mskip\pFqmuskip}
\newcommand{\RpRs}{R_{p}/R_{s}}
\newcommand{\Ngamma}{N_{\gamma}}
\title[Chromatic covariance of scintillation]{Using chromatic covariance to correct for scintillation noise in ground-based spectrophotometry}

\author[J.E. Williams and N.P. Konidaris]{
Jason E. Williams,$^{1,2}$\thanks{E-mail: jewilliams@carnegiescience.edu}, Nicholas P. Konidaris,$^{2}$
\\
$^{1}$Department of Physics and Astronomy, University of Southern California, Los Angeles, CA 90089, USA\\
$^{2}$The Observatories of the Carnegie Institution for Science, 813 Santa Barbara St., Pasadena, CA 91101, USA
\\
}

\date{Accepted XXX. Received YYY; in original form ZZZ}

\pubyear{2023}

\begin{document}
\label{firstpage}
\pagerange{\pageref{firstpage}--\pageref{lastpage}}
\maketitle

\begin{abstract}
Atmospheric scintillation is one of the largest sources of error in ground-based spectrophotometry, reducing the precision of astrophysical signals extracted from the time-series of bright objects to that of much fainter objects. Relative to the fundamental Poisson noise, scintillation is not effectively reduced by observing with larger telescopes, and alternative solutions are needed to maximize the spectrophotometric precision of large telescopes. If the chromatic covariance of the scintillation is known, it can be used to reduce the scintillation noise in spectrophotometry. This paper derives analytical solutions for the chromatic covariance of stellar scintillation on a large telescope for a given atmospheric turbulence profile, wind speed, wind direction, and airmass at optical/near-infrared wavelengths. To demonstrate how scintillation noise is isolated, scintillation-limited exoplanet transit spectroscopy is simulated. Then, a procedure is developed to remove scintillation noise and produce Poisson-noise limited light curves. The efficacy and limits of this technique will be tested with on sky observations of a new, high spectrophotometric precision, low resolution spectrograph.
\end{abstract}

%As evidence, $\pi$ Men c, a \textit{TESS}-confirmed planet that orbits a $V = 5.65$ star, has its spectrophotometric precision reduced to that of a $V = 9$ star due to scintillation.


%%%%%%%%%%%%%%%%%%%%%%%%%%%%%%%%%%%%%%%%%%%%%%%%%%

%%%%%%%%%%%%%%%%% BODY OF PAPER %%%%%%%%%%%%%%%%%%

\section{Introduction}

High-precision spectrophotometry is crucial for a variety of astronomical science cases such as exoplanet transit/eclipse spectroscopy and astroseismology. They all rely on observations of bright stars, where atmospheric scintillation limits the precision of the signals extracted in each of these applications. Atmospheric scintillation refers to the intensity variations that occur from the spatio-temporal refractive index fluctuations caused by atmospheric turbulence. Wavefronts incident on Earth's atmosphere are altered by refractive index fluctuations and intensity variations across the wavefront develop as it travels towards Earth's surface. As the wind drags different patches of turbulence across the line of sight, these intensity variations will change over time, leading to noise\footnote{In this paper, fractional noise is referred to as noise, defined as the inverse of the signal-to-noise ratio, $N/S$.} on the order of $\sim 10$ to $10000$ parts-per-million (depending on exposure time, aperture size, air mass, etc.).
\\

\begin{figure}
 \includegraphics[width=\columnwidth]{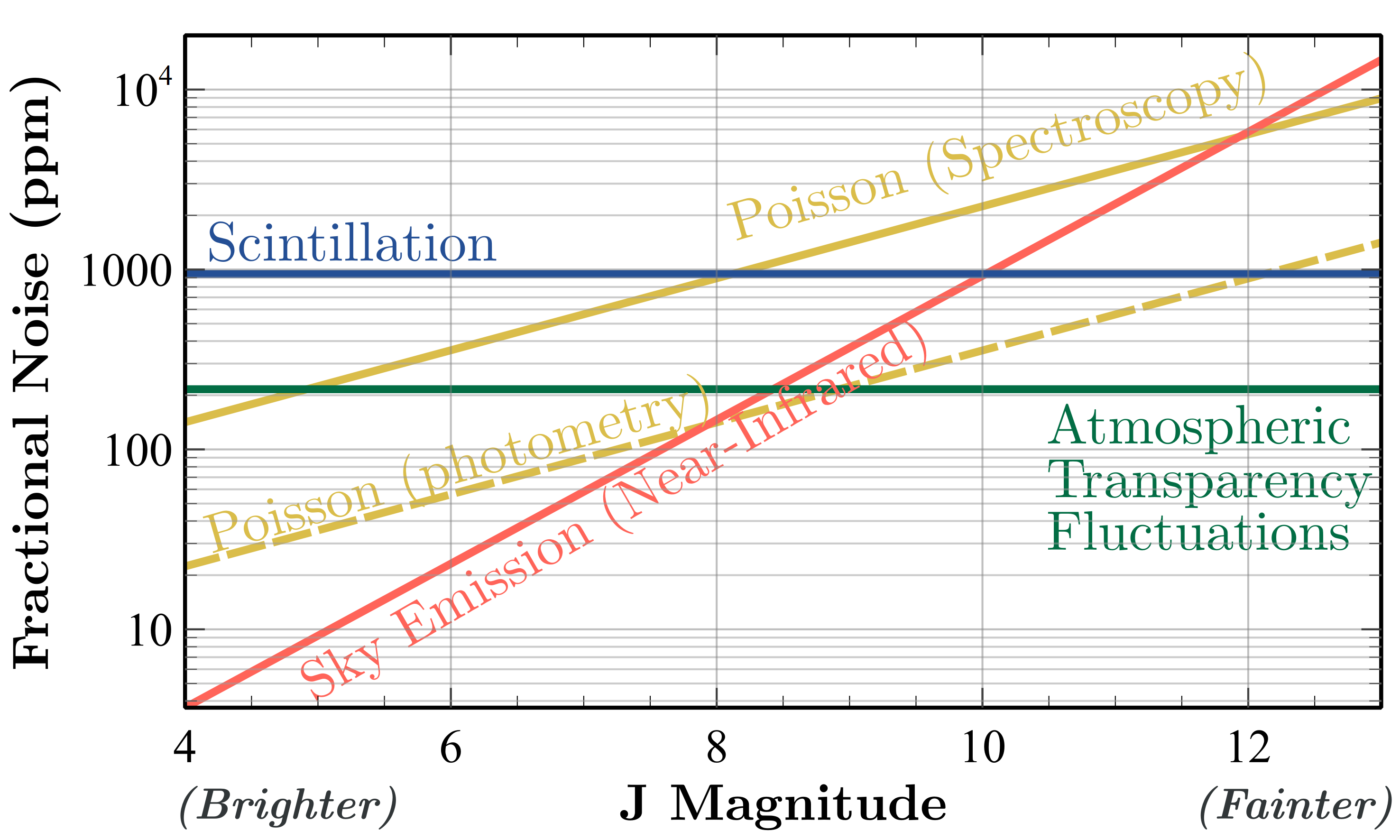}
 \caption{\textbf{Illustration of a ground-based (spectro)photometric noise budget.} As a general rule, scintillation dominates the (spectro)photometric noise budgets in ground-based observations. Since it averages down slower than the photon noise ($D^{-2/3}$ vs. $D^{-1}$), scintillation will always prevent large telescopes from acting as true 'light buckets.'}
 \label{fig:noisebudget}
\end{figure}

As seen in Figure \ref{fig:noisebudget}, one of scintillation's key characteristics is that it does not depend on the number of photons, and therefore dominates the spectrophotometric error budgets of bright objects where the Poisson noise is low. This has the effect of reducing the expected spectrophotometric precision (inverse of the Poisson noise) of bright objects to that of a much fainter object. For instance, despite more than $30\%$ of all \textit{TESS}-confirmed planets having hosts that are $V \leq 9$ (the brightest being $V = 5.65$), their spectrophotometric precisions are limited to $V = 9$ due to scintillation\footnote{Assuming a spectral resolution of $100$ and factoring in an additional $\sqrt{2}$ for differential spectrophotometry.}. This situation does not improve as one increases telescope diameter or exposure time; although the amplitude of scintillation noise will decrease for an increase in exposure time or telescope diameter, bright stars will still have the same precisions as fainter stars. This is shown Over the years, researchers \citep{Gilliland1993, LopezMorales2006, Stefansson2017a, Bryant2020, OBrien2021} have demonstrated scintillation-limited photometry by suppressing the remaining atmospheric and instrumental contributions to the noise budget. This suggests that scintillation noise is likely to be the last hurdle to reaching the fundamental Poisson noise limit for bright stars. 
\\

Many solutions have been proposed to reduce scintillation \citep{Dravins1998, Ryan1998, Kenyon2006,Osborn2011, Hartley2022}, most of which require dedicated instrumentation. If the chromatic covariance of scintillation is accurately known, spectrophotometry can be used to model scintillation as a wavelength-dependent source of noise. The correlation of scintillation is a widely known fact \citep{Tatarski1959, Young1974, Dravins1997a, Kornilov2011a, DeMooij2014, Osborn2015, Limbach2020}, yet few attempts have been made to suppress scintillation in this way.  
\\

This paper provides computationally efficient and accurate analytic formulations of the scintillation covariance at two different wavelengths for a given airmass, telescope diameter, exposure time, and wind velocity. This allows scintillation noise to be accurately isolated from the astrophysical signal and any other sources of noise. In Appendix \ref{sec:scintillationderivationappendix}, a brief overview of the general scintillation equations are given and the scintillation variance is calculated for a general turbulence profile. Section \ref{sec:scintillationdispersion} derives analytic expressions for the scintillation covariance at short and long exposures for given any vertical turbulence profile, wind speed, wind direction, and zenith angle. In the final section, the scintillation covariance is used to simulate ground-based exoplanet transit spectroscopy of a bright star. A general procedure is developed to isolate and remove the scintillation signal from the transit and produce Poisson-noise limited spectrophotometry.

\section{Chromatic covariance of scintillation in ground-based astronomy}
\label{sec:scintillationdispersion}

For bright targets, scintillation noise constitutes a large portion of $\sigma_{\Ngamma}$. During long exposures typical in ground-based spectrophotometry, a common approximation to the scintillation noise comes from \cite{Young1967}

\begin{equation}
\label{equation:youngsapproximation}
    \sigma_{S}^2 \sim 10^{-5} D^{-4/3} \tau^{-1} \sec(\xi)^{3} \exp(-2\,h_{\text{obs}}/H)
\end{equation}

where $D$ is the diameter of the telescope aperture in meters, $\tau$ is the exposure time over which the intensity is recorded in seconds, $\xi$ is the zenith angle, $h_{\text{obs}}$ is the altitude of the observatory and $H$ is the scale height of the atmospheric turbulence. 
This is an approximation in the sense that it recovers all the dependencies on zenith angle, exposure time, and telescope diameter but the scaling factor changes depending on the vertical distribution of turbulence, wind speed, wind direction, etc. For the remainder of this paper, noise will always refer to the fractional noise.\footnote{Fractional means that the deviation is measured relative to the amount of signal, which is the number of photons.} 
The fractional Poisson noise, $\sigma_{p}^{2}$, is equal to the inverse of number of photons detected and is therefore proportional to $\frac{1}{\tau \Delta \lambda D^{2}}$, where $\Delta \lambda$ is the bandpass. The Poisson noise represents the fundamental noise limit on all spectrophotometry. Comparing expressions for the scintillation noise and Poisson noise, several points illustrate why it has been challenging to achieve Poisson noise limited observations for bright objects:
\begin{itemize}
    \item The scintillation noise does not depend on the intensity/number of photons, whereas the Poisson noise decreases with the number of photons.
    \item The ratio of scintillation noise to Poisson noise in independent of exposure and is \textit{weakly} dependent on telescope diameter.
    \item Scintillation noise is independent of the bandpass $\Delta \lambda$.
\end{itemize}

The first two points mean that when scintillation dominates over Poisson noise the ratio between the two cannot be decreased (substantially) by increasing the telescope size or exposure time. The last point is an interesting one. If scintillation is independent of $\Delta \lambda$ this suggests that scintillation cannot be decreased (or increased) by observing at different wavelengths or by combining different wavelengths. Therefore, scintillation is essentially achromatic and observations at multiple wavelengths may be used to remove scintillation noise. As mentioned in the previous section, this fact is well known. What is not well known is the exact form of the scintillation covariance given any set of observational parameters. Without an accurate model of the covariance, any attempt to isolate the scintillation noise may perturb the underlying signal. The covariance of the scintillation noise at two different wavelengths primarily depends on 1) how similar the paths travelled by the two chromatic wavefronts are and 2) if the aperture is used to observe stellar scintillation is larger than the Fresnel radius\footnote{The Fresnel radius, $r_F$, is the length scale of turbulence for which diffractive intensity variations begin to develop across the wavefront after some propagation distance $H$ and some wavelength $\lambda$. It is defined as $r_F := \sqrt{\lambda H}$. For visible wavelengths and long path lengths ($\sim 10 \,$\text{km}), $r_{F} \sim 20 \,\text{cm}$.} \citep{Roddier1981}. Therefore, spectrophotometry at similar wavelengths (which travel through similar regions of atmospheric turbulence) observed using large telescopes will experience a high degree of correlation. One of the first known explorations of this phenomena was in \cite{Tatarski1959}, who found good agreement between their analytical predictions and observations. Since their aperture size was similar to the Fresnel radius for those observations, the averaging effect of the telescope size could be ignored in those calculations. This section seeks to extend this work, developing a formalism to calculate the scintillation covariance between two wavelengths on a large telescope for a given zenith angle, wind speed, and wind direction. 

\subsection{Expressions for the scintillation covariance}
\label{ssec:groundbasedscintillationcovariance}
The most general expression for the scintillation covariance of two monochromatic wavefronts with wavelengths $\lambda_{1}$ and $\lambda_{2}$ passing through different turbulent regions of the atmosphere is given in the first equation in \cite{Hill1989}. For case for stellar scintillation (i.e. plane waves) observed on a single telescope, the scintillation covariance simplifies to 

\begin{multline}
    \label{equation:scintillation_dispersion_covariance_shortexp}
    \sigma_{\lambda_{1}, \lambda_{2}}^{2} = \frac{9.62\, }{\lambda_1  \lambda_2} \int_0^{\infty} df \, f \Phi(f) \, \int_0^{H/\sec(\xi)} dz_{\xi} \, C_{n}^2(z_{\xi}) \\[7pt]
    \qquad \times \, \sin(\pi \lambda_1 z_{\xi} f^2) \, \sin(\pi \lambda_2 z_{\xi} f^2) \left(\frac{2 J_{1}(\pi D f)}{(\pi D f)}\right)^2 J_0(2\pi f \rho(z))
\end{multline}

\noindent where  $\lambda_1$ and $\lambda_2$ are the wavelengths of each wavefront, $f$ is the spatial frequency of refractive index fluctuations, $z$ is the altitude, $\xi$ is the zenith angle, $z_{\xi}$ is defined to be $z\sec(\xi)$, $D$ is the receiving aperture diameter, and $C_{n}^{2}(z)$ is the vertical distribution of refractive index fluctuations (referred to as the turbulence profile). $\Phi(f)$ is power spectrum of refractive index fluctuations, assumed to be a Kolmogorov spectrum equal to $f^{-11/3}$, $\rho$ is the separation between the two monochromatic waves along the line of sight calculated in Equation \ref{equation:pathseparation} and $J_0$ is the zeroth order spherical Bessel function \citep{Abramowitz1965}. The addition of $J_0$ arises due to the shift property of Fourier transform \citep{Kornilov2012}. A relative shift in the spatial coordinates of the refractive index fluctuations corresponds to multiplication of the spatial frequency power spectrum by $e^{2\pi i \vec{f} \cdot \vec{\rho}}$. Evaluating this in spherical coordinates and then integrating over all polar angles gives the Bessel function. For a flat isothermal atmosphere, the line-of-sight displacement between two wavefronts at wavelengths $\lambda_1$ and $\lambda_2$ at some altitude $z$ is \citep{Caccia1988}

\begin{multline}
    \label{equation:pathseparation}
    \rho(\lambda_1, \lambda_2, \xi, z) = \text{tan}(\xi)\,\text{sec}(\xi)\,\left[n(\lambda_1) - n(\lambda_2)\right] \\[7pt]
    \times \left(e^{-\frac{h_0}{H_0}}(H_0 - (H_0 +  z)e^{-\frac{z}{H_0}}\right)
\end{multline} 

\noindent where $h_0$ is the observatory height and $H_0$ is the scale height of the atmosphere ($\sim 8300$m). From Equation \ref{equation:scintillation_dispersion_covariance_shortexp}, the aperture averaging approximation (Equation \ref{equation:lowfrequencyapproximation}) is made along with the same trigonometric identity used in Appendix \ref{sec:scintillationvariance} to arrive at

\begin{multline}
    \label{equation:scintillation_dispersion_covariance_shortexp3}
     \sigma_{\lambda_{1}, \lambda_{2}}^{2}(\rho) \approx \frac{0.32\, \Gamma(-5/6)}{4\lambda_1  \lambda_2} \int^\infty_0 dz_{\xi}\, C_n^2(z_{\xi}) \\[7pt] \times \mathfrak{Re} \, \Big[G_{-}(z_{\xi}\,{;}\, x^+) - G_{+}(z_{\xi}\,{;}\, x^{-}) \Big]
\end{multline}

\noindent where 

\begin{equation}
G_{\pm}(z_{\xi}\,{;}\, x^{\pm}) = L_{5/6}(-x^{\pm}) \left((bD)^2 - 2\pi i\,z_{\xi} \lambda_2 (1 \pm \frac{\lambda_1}{\lambda_2}) \right)^{5/6}
\end{equation}

\noindent This is almost identical to the expression derived in Equation \ref{equation:scintillationcovariancepowerspectrum4}, with the path separation factored in via the Laguerre function. Note that $b$ here is an empirical parameter $~ 2.7$, see \ref{sec:scintillationderivationappendix} for more details.
\\

\par
\begin{figure}
 \includegraphics[width=\columnwidth]{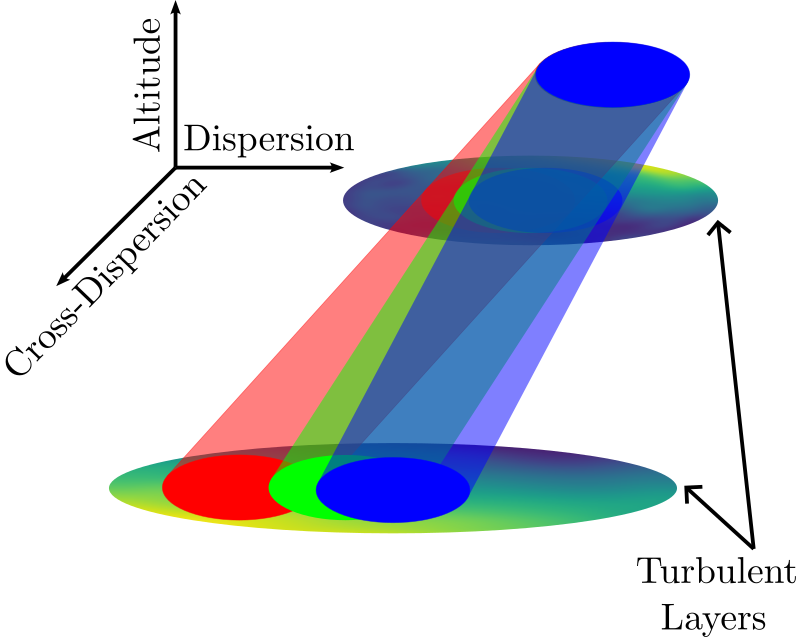}
 \caption{\textbf{Illustration of dispersed wavefronts passing through layers of atmospheric turbulence.} At the top of the atmosphere, all chromatic starlight enters undispersed. On the way from the top of the atmosphere to the telescope, chromatic wavefronts are dispersed and travel through layers of atmospheric turbulence. The direction of atmospheric dispersion (here, left to right on the page) introduces a preferential direction for the wind speed. When the wind blows in the direction of atmospheric dispersion, all wavefronts see nearly the exact same turbulence. However, when the wind blows perpendicular to the direction of atmospheric dispersion (cross-dispersed direction, here in/out of the page), there are always portions of each wavefront that are exposed to different regions of turbulence. This leads to the wavelength correlation of scintillation depending on the direction of the wind speed relative to the direction of atmospheric dispersion.}
 \label{fig:dispersionillustration}
\end{figure}

 In comparison to the short-exposure regime, the behavior of the long-exposure covariance is determined by the relationship between the telescope diameter, exposure time, the direction of dispersion, and wind speed. The illustration in Figure \ref{fig:dispersionillustration} helps to show how these two are related. For parallel wind speeds (oriented along the direction of dispersion), the atmospheric turbulence is blown in the direction of each monochromatic wavefront, and therefore as time goes on the integrated turbulence seen by each monochromatic wavefront becomes more and more similar. For winds that blow turbulence orthogonal to the direction of atmospheric dispersion, the opposite is true - the exposure time has less of an effect since the atmospheric turbulence is , since the regions of turbulence that are common-mode and differential-mode remain constant over time (relative  and averaged regions of turbulence remain constant with exposure time. Note that in both cases, the total variance decreases with exposure time, independent of the wind direction (see \ref{sec:scintillationvariance}). For an exposure time $\tau$, wind speed $w$, and an angle $\theta$ between the wind speed and the direction of dispersion, the bichromatic covariance of scintillation is given by 

\begin{multline}
    \label{equation:scintillation_dispersion_covariance_longexp}
    \sigma_{\lambda_1, \lambda_2}^{2}(\tau) =  \sec(\xi) \int_{0}^{H} dz \, C_{n}^{2}(z)\, Y(z,\lambda_1, \lambda_2, \tau, w(z), \theta(z))
\end{multline}

\noindent where 

\begin{multline}
    \label{equation:scintillation_dispersion_covariance_longexp1}
    Y(z,\lambda_1, \lambda_2, \tau, w(z), \theta(z)) = \frac{0.32}{\lambda_1  \lambda_2} \int_0^{\infty} df  \, f^{-8/3} J_0(2\pi f \rho) \\[7pt] \qquad  A_{w}(...) \, \sin(\pi \lambda_1 z_{\xi} f^2) \, \sin(\pi \lambda_2 z_{\xi} f^2) \left(\frac{2 J_{1}(\pi D f)}{(\pi D f)}\right)^2
\end{multline}

The function $A_{w}$ is the wind-averaging filter and is defined using the angle formalism from \cite{Kornilov2011a} as

\begin{equation}
    \label{equation:sincsquared1}
    A_{w}(...) = \frac{1}{2\pi}\int_0^{2\pi} d\phi \, e^{2\pi i f \rho \cos(\phi - \theta)} \text{sinc}^2(w\tau f \cos(\phi)) 
\end{equation}

\noindent where $\theta$ is the angle between the direction of dispersion and the direction of the wind speed. In its present form, no analytical solution could be found in integral tables. The key to solving this integral is to convert the exponential into its series form

\begin{equation}
    \label{equation:windaveragingfilter_parallel}
    A_{w}(...) = \sum_{n = 0}^{\infty} \frac{(2\pi i f \rho)^n}{2\pi\,n!} \int_{0}^\infty d\phi \, \cos^n(\theta - \phi) \, \text{sinc}^2(w\tau f \cos(\phi)) 
\end{equation}

and to rewrite the $\text{sinc}^2$ term as 

\begin{equation}
    \text{sinc}^2(w \tau f \cos(\phi)) = \frac{1}{w^2 \tau^2 f \cos(\phi)}\int_0^{w\tau} dt \, \sin(2 f t \cos(\phi))
\end{equation}

The details can be found in the Appendix. The final result is 

\begin{multline}
\label{equation:final_long_exposure_covariance}
    \sigma_{\lambda_1, \lambda_2}^{2}(\tau) \approx \frac{0.32}{16\pi\lambda_1  \lambda_2} \int_{0}^{H/\sec(\xi)} dz_\xi \, C_{n}^{2}(z_{\xi}) \\[7pt] \qquad \sum_{n = 0}^{\infty} \frac{(2\pi i \rho)^n}{(w^2 \tau^2 n!)} \sum_{k=0}^{n} G(n,k,\theta) \frac{\Gamma(\frac{k+1}{2})\,\Gamma(\frac{n-k-1}{2}) \, \Gamma(\frac{n}{2} - \frac{11}{6})}{\Gamma(\frac{n}{2})} \\[7pt] 
    \times \mathfrak{Re} \left[F_{n}^{-}(\tau, w, z_{\xi}) - F_{n}^{+}(\tau, w, z_{\xi}) \right]
\end{multline}

\noindent where 
\begin{multline}
\label{equation:hypergeom}
F_{n}^{\pm}(\tau, w, z_{\xi}) = \left(b^2 D^2 - i\pi \lambda_2 \left(1 \pm \frac{\lambda_1}{\lambda_2}\right) z_{\xi}\right)^{11/6 - \frac{n}{2}} \\[7pt]{}_{2}F_{2}\left[\left\{\frac{n}{2} - \frac{11}{6}, \frac{n-k-1}{2}\right\},\left\{\frac{1}{2}, \frac{n}{2}\right\},{\frac{-w^2 \tau^2}{b^2 D^2 - i \pi \lambda_{2} (1 \pm \frac{\lambda_1}{\lambda_2}) z_{\xi}}}-1\right]
\end{multline}
\\

\noindent Therefore the weighting function is

\begin{multline}
\label{equation:longexposure_covariantweightingfunction_final}
Y(...) = \frac{0.32 \, \sec(\xi) \, \mathfrak{Re} \left[F_{n}^{-}(\tau, w, z_{\xi}) - F_{n}^{+}(\tau, w, z_{\xi}) \right]}{16\pi\lambda_1  \lambda_2}  \\[7pt] \times \sum_{n = 0}^{\infty} \frac{(2\pi i \rho)^n}{(w^2 \tau^2 n!)} \sum_{k=0}^{n} G(n,k,\theta) \frac{\Gamma(\frac{k+1}{2})\,\Gamma(\frac{n-k-1}{2}) \, \Gamma(\frac{n}{2} - \frac{11}{6})}{\Gamma(\frac{n}{2})} 
\end{multline}
\\

\begin{figure}
 \centering
 \includegraphics[width=3in]{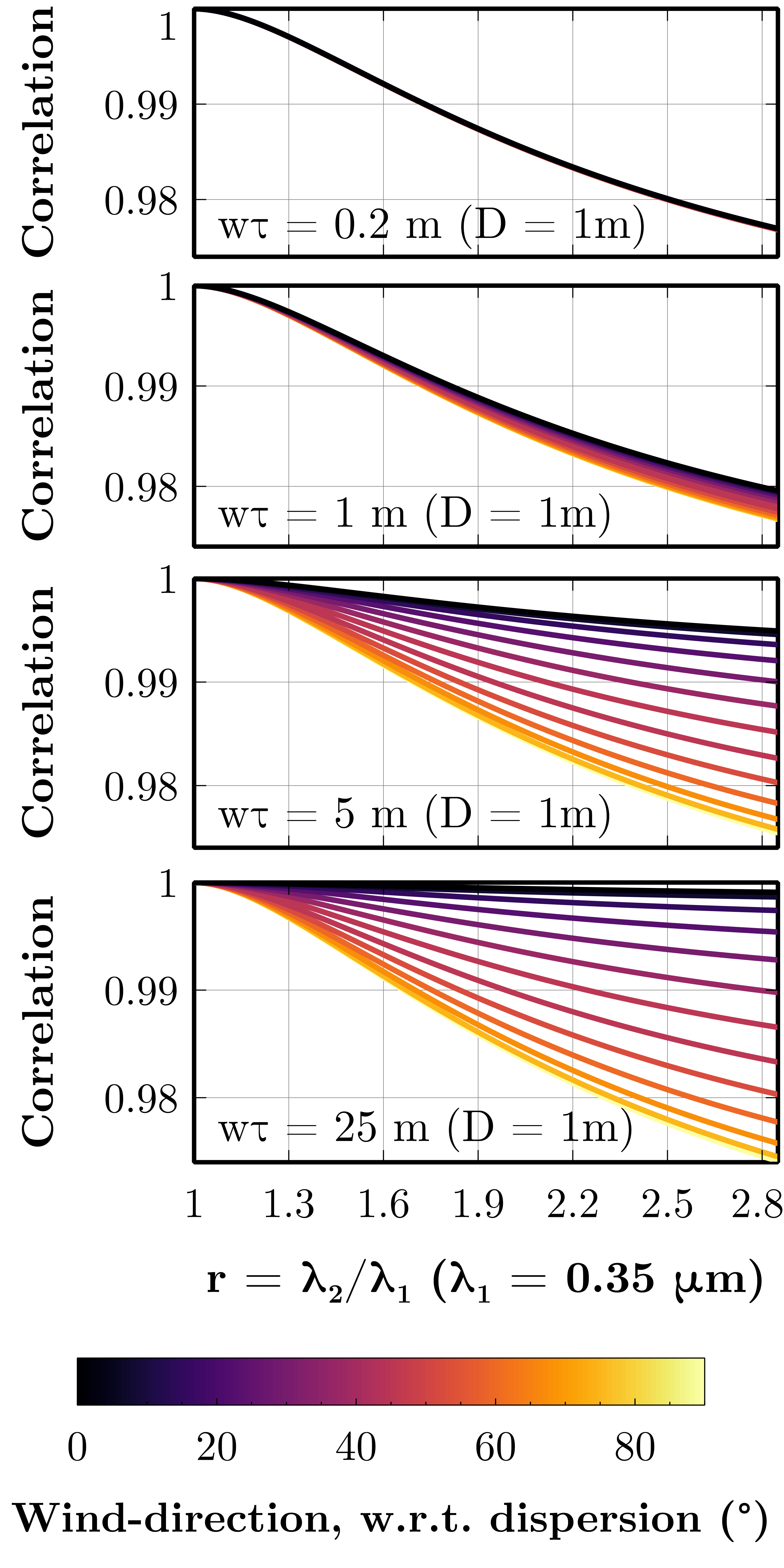}
 \caption{\textbf{How the covariance changes with exposure time and wind speed.} For $w\tau < D$, the wind-direction does not influence the correlation between two wavefronts. This is because, when averaged over the telescope area, the area of turbulence that is common/normal to each wavefronts is the same. As $w \tau$ approaches and surpasses the telescope diameter, that the wind direction begins to play a role in the correlation coefficient. For completely perpendicular winds, the disjoint area grows as $\approx \rho w \tau$ whereas for parallel winds, the area is constant with exposure time.}
 \label{fig:covariance_with_angle}
\end{figure}

\begin{center}
\begin{figure}
 \centering
 \includegraphics[width=3in]{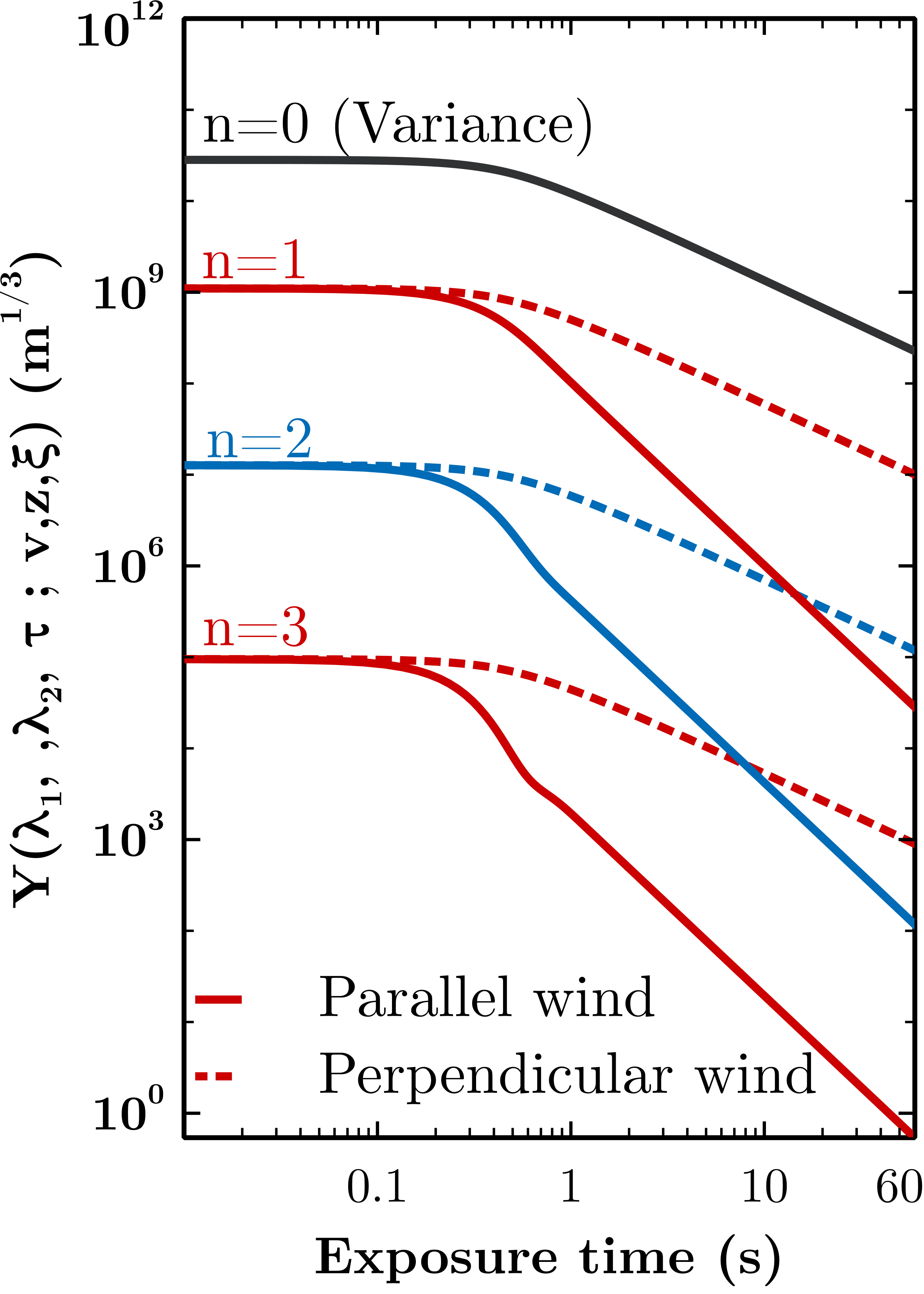}
 \caption{\textbf{Comparing the parallel/perpendicular wind covariance weighting functions as a function of exposure time.} The first $n=4$ terms of covariant weighting function, Equation \ref{equation:longexposure_covariantweightingfunction_final} are shown. The covariant weighting function is found by summing all terms together. The red curves represent the subtracted terms, and the blue curve represents the additive term.  The covariant weighting function for perpendicular winds has a long-exposure slope equal to the variance/$n=0$ weighting function long-exposure slope. For parallel winds, the long-exposure covariance has a stronger slope leading to a larger covariance as the exposure time increases.  }
 \label{fig:variance_vs_covariance_exposuretime}
\end{figure}
\end{center}

\noindent The key behavior of the weighting function $Y(z, \lambda_1, \lambda_2, \tau, w, \theta)$ for a given altitude is shown in Figure \ref{fig:covariance_with_angle}. Rather than plotting the absolute value, a kind of correlation is plotted - $Y(z, \lambda_1, \lambda_2, \tau, w, \theta) / Y(z, \lambda_1, \lambda_1, \tau, w, \theta)$. The primary behavior is controlled by the quantity $w \tau$, which is effectively the area of turbulence that is averaged during an exposure, the telescope aperture $D$, and the angle between the direction of atmospheric dispersion and the wind speed, $\theta$. For $w\tau < D$, the wind direction does not influence the correlation between two wavefronts. This is because, when averaged over the telescope area, the area of turbulence that is common/normal to each wavefront is the same. As $w \tau$ approaches and surpasses the diameter of the telescope, the wind direction begins to play a role in the correlation coefficient. For wind directions that are normal to the direction of dispersion, the non-overlapping area grows as $\approx \rho w \tau$, while for parallel winds, the area is constant with exposure time. This is why for a wind-direction of $90\deg$ with respect to atmospheric dispersion, the correlation does not change despite increasing the exposure time or wind speed. The source of this behavior can be further illustrated by plotting the absolute value of the weighting function. In Figure \ref{fig:variance_vs_covariance_exposuretime}, which shows how the terms in the parallel/perpendicular wind weighting functions change as a function of exposure time, leading to greater covariance for parallel winds and less covariance for perpendicular winds.
%Remember to FIX zenith dependence of velocity!!
%Fix minus signs
%fix wolfram referencesAZQ
%fix figure 3 shows correlation not weighting function
\\

\section{Isolating Scintillation noise with Spectrophotometry}
In the temporal domain and with long exposures, behaves as white noise \citep{Dravins1998,Kenyon2006,Fohring2019} - there is no correlation between exposures. Since scintillation is correlated in wavelength space, knowledge of the chromatic covariance of scintillation enables multiwavelength time series to be identified separately from astrophysical signals (which may also vary with wavelength). In theory, this paves the way for reaching the Poisson noise limit and obtaining more precise measurements of astrophysical parameters \citep{Stefansson2017a}. To demonstrate this, a realistic simulation of ground-based spectrophotometry is developed using the equations developed in the previous section. The astrophysical signal to be recovered is the transmission spectrum of an exoplanet atmosphere. Two separate observations are simulated to simulate the realistic scenario of combining data from two different instruments, or on two separate nights. The details of the simulation are detailed in Appendix \ref{ssec: simulating_ground_based_tspec}. In addition to the assumptions in the simulation outlined in the Appendix, there are a number of additional assumptions that were made during this simulation. They are summarized below:
\begin{itemize}
	\item \textit{No sources of time-correlated noise.} Time-correlated noise is known to affect ground-based spectrophotometry through variable atmospheric absorption and turbulence and its complicated interactions with instrumentation (i.e. time-dependent aperture losses, flat-field imperfections, etc.). In order to isolate scintillation noise using the coming method, time-correlated noise must be kept sufficiently low (well-below the Poisson noise) or its covariance must be accurately known.
	\item \textit{Kolmogorov power spectrum with no outer scale.}
For large telescopes which suppress (average over) high frequency fluctuations, taking the upper limit of the frequency to infinity is not an issue. The shape of the power spectrum $\Phi(f)$ in the low frequency range is important for large telescopes and for long exposures. This simulation assumes a Kolmogorov spectrum, which has the effect of overestimating the scintillation variance/covariance compared to models that factor in an outer scale. This is a safe assumption for telescopes with diameters smaller than 5m \citep{Osborn2015a}.
 
	\item \textit{Effects of central obstruction and wavelength averaging play a minor role.} For long exposures on large telescopes, the central obstruction plays a minor role and is therefore not factored into any equation in the previous sections \citep{Osborn2015}. 
	
	\item \textit{No wavelength averaging.} Each wavelength bin in the scintillation simulation is simulated as a monochromatic wavelength and therefore the effects of wavelength averaging are assumed to play a minor role due to the near achromatic nature of scintillation on large telescopes \citep{Tokovinin2003}.
	
	\item \textit{The zenith dependence of the wind speed.} The projected wind speed across a turbulence layer has a known zenith and directional dependence as identified by \citep{Young1969}. When the direction of the wind speed and the direction of dispersion are codirectional, the projected speed is $v_{\text{proj}} = v/\text{sec}(\xi)$. When the two directions are orthogonal, the projected velocity equals the wind speed. This has the effect of reducing the covariance between wavelengths at large zenith angles. Since most astronomical observations tend to agree with the $\text{sec}(\xi)^3$ dependence of scintillation, the zenith dependence of the wind speed is omitted for simplicity.
	
	\item \textit{Simple picture of atmospheric refraction.}
This simulation assumes a relatively simple picture of atmospheric refraction. First, implicit in Equation \ref{equation:pathseparation} is that Earth's atmosphere is isothermal, which it is not. Therefore the path separation between given by Equation \ref{equation:pathseparation} is likely to be inaccurate for regions in the lower atmosphere where the pressure-temperature relationship is not constant. Secondly, the strength of refractive index fluctuations has an additional wavelength dependence due to the wavelength dependence of the refractive index. This effect is assumed to have a small effect on the overall (co)variance and is therefore not simulated.

\end{itemize}

\begin{figure}
 \includegraphics[width=\columnwidth]{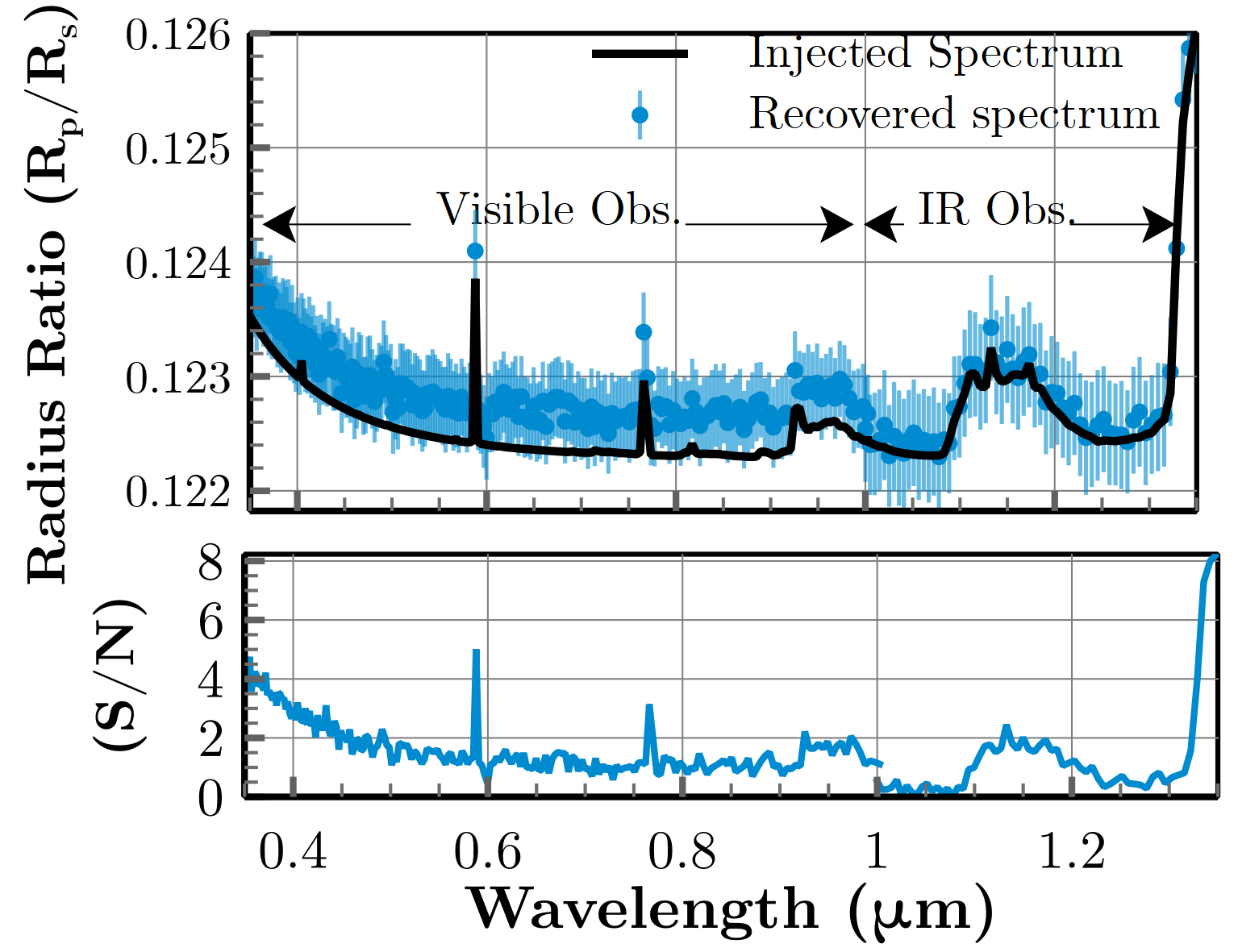}
 %add cn2 profile
 \caption{\textbf{Injected transmission spectra along with the recovered spectra prior to scintillation correction.} Since scintillation is a nearly achromatic stochastic noise source, this adds the same bias to each spectrophotometric light curve, leading to a bias in the recovered transit spectra. This also causes the recovered radius ratio errors to appear inflated, since each light curve is fit independently without any knowledge of the covariance.}
 \label{fig:final_spectrum_before}
\end{figure}

\subsection{Identifying scintillation noise using its chromatic covariance}

The resulting transmission spectrum from the simulation detailed in Appendix \ref{ssec: simulating_ground_based_tspec} is shown in Figure \ref{fig:final_spectrum_before}. The signal-to-noise is also plotted using Equation (4) from \cite{Phillips2021} (except the minimum radius ratio is used instead of the median radius ratio). Since the injected spectrum is known, the chi-squared (normalized by the degrees of freedom) can be calculated for the recovered spectrum. The $\chi^2$ is $< 1$, which suggests that the data errorbars are likely overestimated. Note, that there is also a bias/offset between the $.35 {\mu}$m - $1{\mu}$m 'visible' observation and the $1{\mu}$m - $1.35{\mu}$m 'infrared' observation. This bias is induced by the presence of scintillation - covariance across wavelengths necessarily leads to a bias in the inference across wavelengths. Offsets like these have been commonly observed in ground-based transmission spectroscopy when combining datasets \cite{McGruder2022}, yet the origin is still unknown.
\\

\begin{figure}
 \includegraphics[width=\columnwidth]{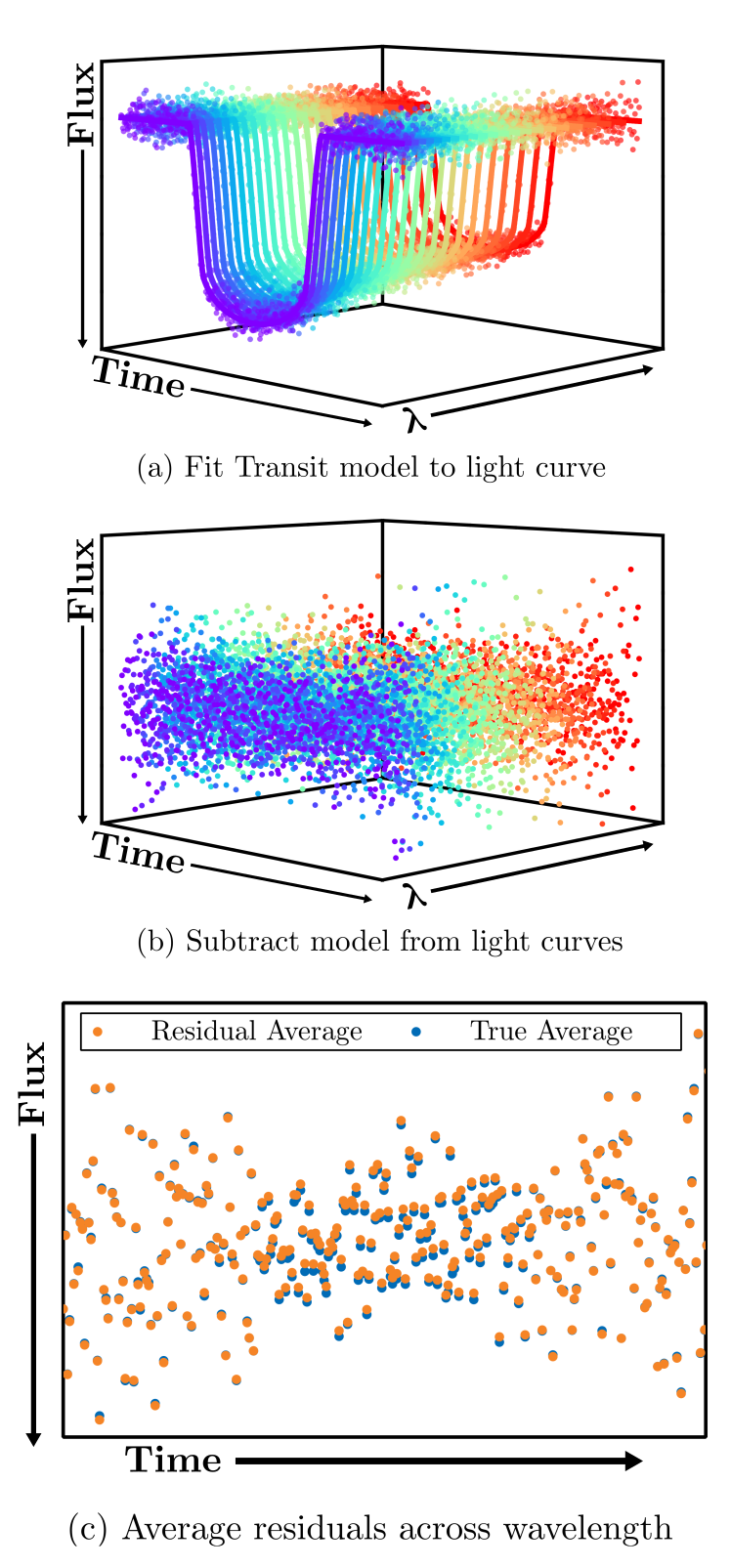}
 %add cn2 profile
 \caption{\textbf{First step in the scintillation suppression recipe.} The first step to removing scintillation is to obtain an accurate estimate of the (nearly) achromatic scintillation noise. To do this, first transit models are fit to each
light curve and then subtracted. Then the resulting residuals are then averaged together in wavelength space. The rolling standard deviation is calculated using these averaged residuals, generating an estimate for the scintillation noise, $\sigma_{s}.$}
 \label{fig:correctionstep1}
\end{figure}
 
The simulated light curves only contain scintillation noise and Poisson noise and are modeled according to 
\begin{equation}
    \label{equ:lightcurve_distribution}
    \textbf{y}_{s+p} \sim \mathcal{N}(\textbf{T}, \mathbf{\Sigma}_{s+p})
\end{equation}

\noindent where $\mathcal{N}$ is the normal distribution, $\textbf{T}$ is the transit model at each wavelength for a set of exposures, and $\mathbf{\Sigma}_{s+p}$ is the covariance matrix in time and wavelength with contributions from scintillation and Poisson noise. $\textbf{T}$ is size $N \times 1 \times M \times 1$ and $\mathbf{\Sigma}_{s+p}$ is size $N \times N \times M \times M$\footnote{In this section, bolded variables refer to quantities that have dimensions greater than $1$ and unbolded variables will refer to quantities with dimensions less than $1$.}, where $N$ is the number of exposures and $M$ is the number of wavelengths. Since scintillation and Poisson noise are uncorrelated, $\mathbf{\Sigma}_{s+p} = \mathbf{\Sigma}_{s} + \mathbf{\Sigma}_{p}$. 
\\

First, an estimate for the scintillation variance at each wavelengths is needed. As mentioned in the previous section, the variance of scintillation is essentially achromatic. If a light curve fit is subtracted for each wavelength channel the residuals will be dominated by scintillation noise and photon noise. Since the scintillation noise is correlated across wavelengths and the Poisson noise is not, averaging these residuals in wavelength space will then leave the scintillation essentially unchanged while simultaneously decreasing the photon noise. If $M$ channels are averaged together, the photon noise decreases as $\sqrt{M}$ while the scintillation noise does not. This provides an accurate estimate of the scintillation noise realized in this observation. To estimate the scintillation noise used to generate this realization, the rolling standard deviation can be calculated from the averaged residuals. To find the right window size for the rolling standard deviation, power spectra of the scintillation noise from 22 nights at Paranal Observatory are averaged \cite{Osborn2018}. The averaged scintillation noise power spectrum has a $\frac{1}{f}$ power spectrum down to a frequency of ~ 2 mHz, and then transitions to a white noise spectrum. Therefore, a window size of 10 minutes should capture the low frequency variations in scintillation noise with some accuracy. 
\\

The first step is illustrated in Figure \ref{fig:correctionstep1}. The source of the bias can be seen in panel $(c)$, where the estimated residual average is offset from the true average for all exposures during the transit. Removing the bias requires the most accurate model of the scintillation covariance, which requires time-dependent knowledge of the altitude dependent wind speed/direction and are typically not quantities available to most observers. Therefore, an approximation to the scintillation covariance needs to be made. Using the estimate for the scintillation noise, $\sigma_{s}(t_{i})$, a time-dependent covariance is generated according to
%change this to tensor product
\begin{equation}
    \Sigma_{s}(t_{i}) \approx \sigma_{s}\, \mathbf{J}
\end{equation}

\noindent where $J$ is the ones matrix of size $M$. This amounts to saying that scintillation is perfectly correlated across all wavelengths at all times. As seen in Figure \ref{fig:covariance_with_angle}, this is a good approximation on large telescopes and when taking long exposures. To complete the full covariance, the Poisson noise must be added to the diagonals. The Poisson noise can easily be estimated during real observations - here it is measured using the standard deviation of the difference between residuals and the averaged residuals (across wavelength space). The final covariance matrix at each time is then approximated as:

\begin{equation}
    \Sigma_{s+p}(t_i) \approx \sigma_{p}\,\mathbf{I} + \sigma_{s}(t_{i})\, \mathbf{J} 
\end{equation}

\noindent This matrix has a closed-form inverse using the Sherman-Morrisson formula. To remove the bias in the inferred radius ratios, a new transit model is generated based on the initial radius ratios $(\overrightarrow{\RpRs})_{0}$, 
\begin{equation}
\label{equation:Transitmodeloffset}
\textbf{T}_{g,s} = F\left((\overrightarrow{\RpRs})_{0} + gs, i, a/R_{s}\right)
\end{equation}

\noindent where $g$ is the radius ratio offset, and $s$ is the sign of the offset. The best values that remove the bias are found by maximizing the likelihood according to 

\begin{equation}
\label{equation:scintillationsuppressedlikelihood1}
\mathcal{L}_{p} \sim | \mathbf{\mathbf{\Sigma}}_{s+p} |^{-1/2} \exp\left[-\left(\frac{1}{2}\right)\, (y-\textbf{T}_{g,s} )^{T} \mathbf{\mathbf{\Sigma}}_{s+p}^{-1} (y-\textbf{T}_{g,s} )\right]
\end{equation} 

\begin{figure}
\includegraphics[width=\columnwidth]{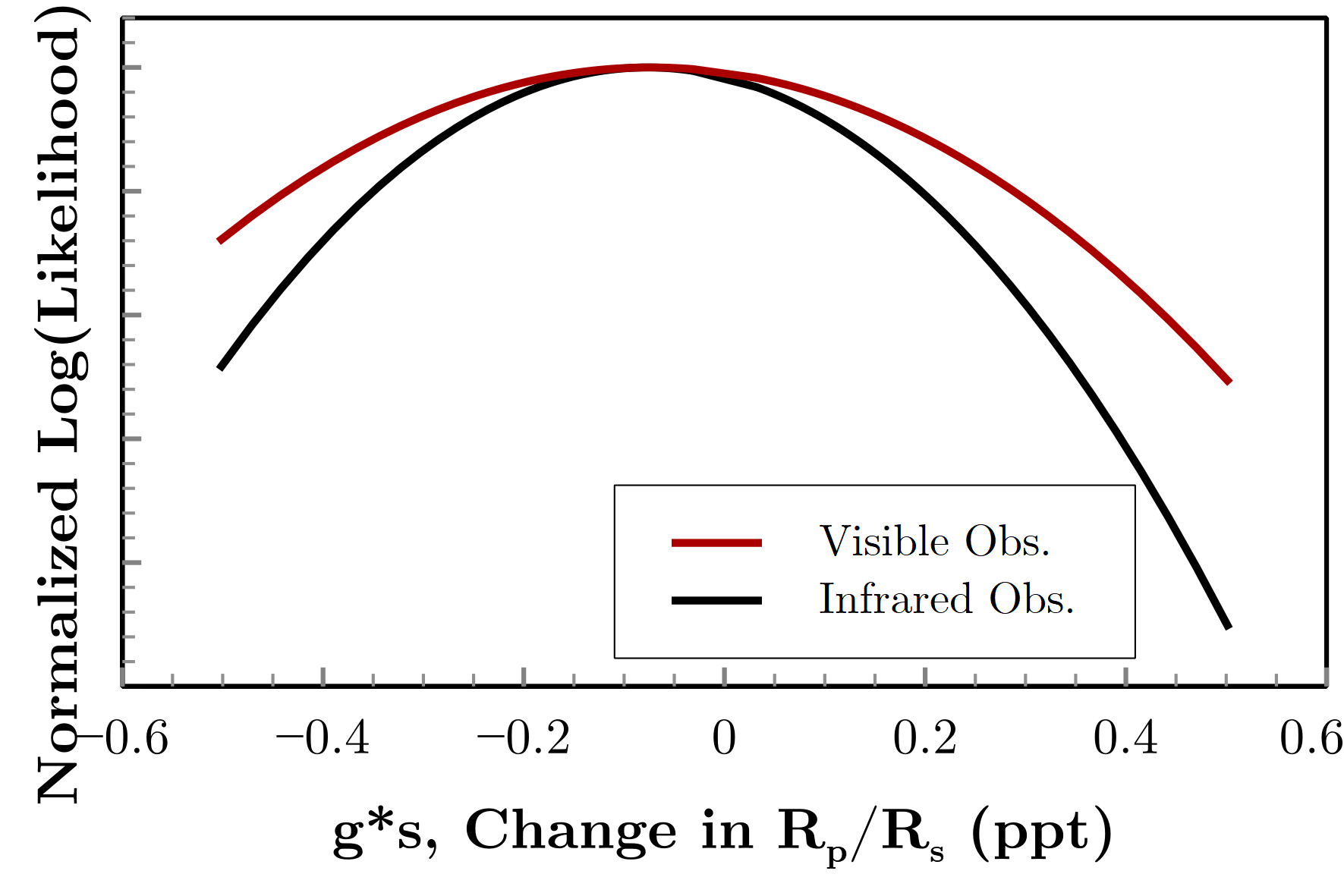}
%add cn2 profile
\caption{\textbf{Removing the scintillation bias.} The scintillation bias comes from the scintillation noise inducing a nearly achromatic signal with respect to wavelength. This manifests itself as a bias in the astrophysical signal with respect to wavelength. The bias in the inferred radius ratio can be removed offsetting the original radius ratio by some amount $g\times s$ and then evaluating the likelihood in Equation \ref{equation:scintillationsuppressedlikelihood1}. The maximum of this likelihood across all offsets is the amount each radius ratio is shifted by to remove the bias.}
\label{fig:loglikelihood}
\end{figure}

Figure \ref{fig:loglikelihood} plots Equation \ref{equation:scintillationsuppressedlikelihood1} for the visible and infrared observations, showing a distinct maximum as a function of $g$.   
\\

\begin{figure}
 \includegraphics[width=\columnwidth]{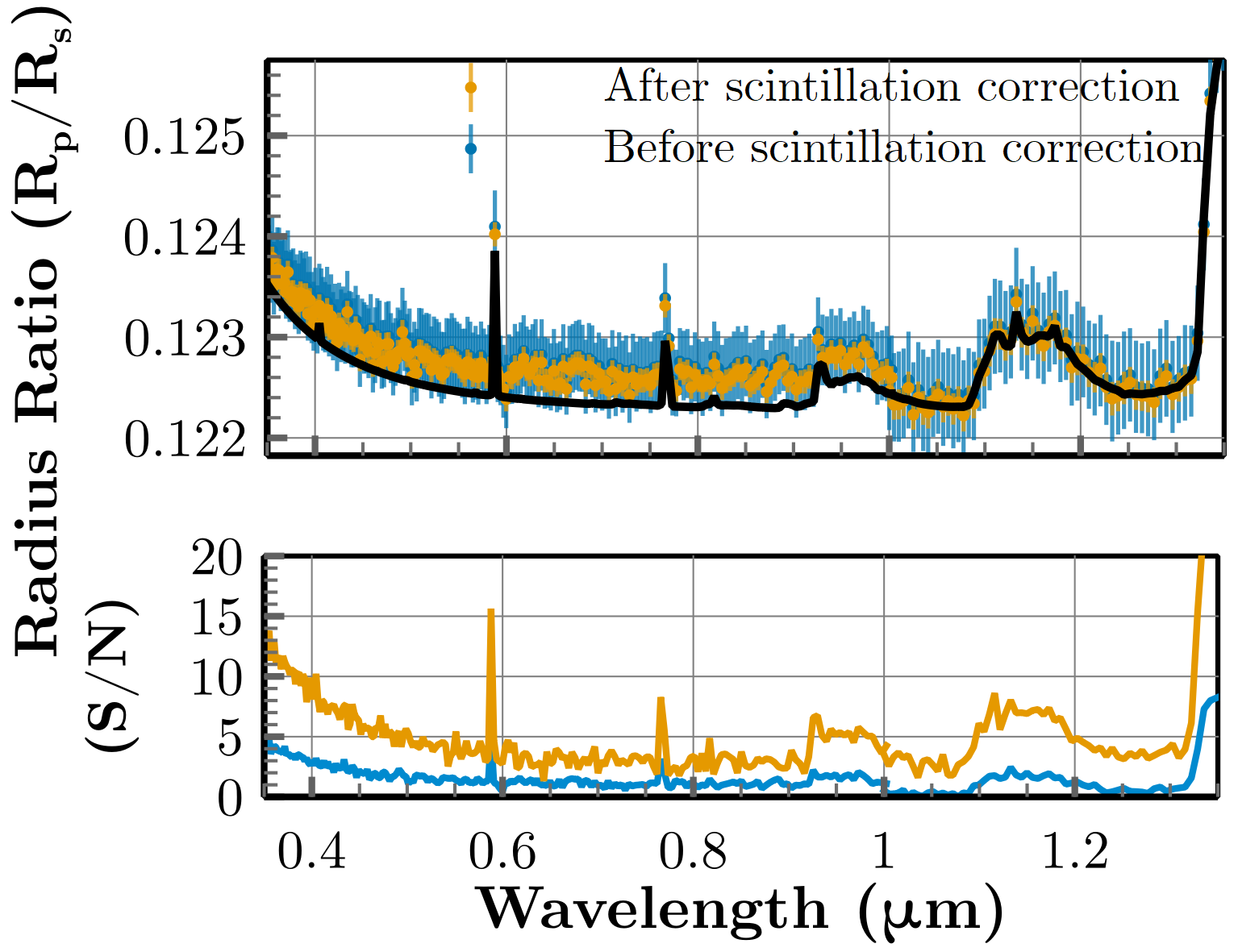}
 \caption{\textbf{Injected transmission spectra along with the recovered spectra prior to scintillation correction.} After incorporating the scintillation correction, the inferred radius ratios are de-biased and accurate error bars are obtained.}
 \label{fig:final_spectrum_combined}
\end{figure}

\noindent Once the new radius ratios are found, new transits can be calculated, new residuals can be calculated and averaged in wavelength space again. This averaged residual can subtracted from each raw light curve to remove scintillation noise, leaving just Poisson noise behind. A final fit of these scintillation-removed light curves will now give accurate error bars. The resulting transmission spectrum is shown in Figure \ref{fig:final_spectrum_combined}. 

\subsection{Discussion}
\label{ssec:Discussion}

In this simulation, the scintillation noise is suppressed to levels well below the Poisson noise, despite being 3$\times$ greater in amplitude. The main limitation to how well scintillation noise can be estimated and then removed is the number of light curves used to average the Poisson noise down. Therefore this technique can work even when light curves have similar amounts of Poisson and scintillation noise, as long as the Poisson noise of the entire bandpass is much less than the scintillation noise. 
\\

This procedure was able to increase the precision on the radius ratio reach the Poisson noise limit, but it was less successful at removing the scintillation induced-bias in the visible than in the infrared. This was a consistent theme throughout multiple different scintillation realizations. The reason for the difference in efficacy is likely because the approximation of the scintillation covariance matrix by a constant is a much better approximation in the infrared because the atmospheric dispersion is less. One solution to this was to was to perform the scintillation bias correction on smaller regions of the bandpass, where the covariance can remain much higher between wavelengths. However, this was still inconsistent in its accuracy. The more accurate solution was to generate a time-series of best-fit values for the altitude $z$, wind speed $w$ and wind direction $\theta$ to model how the covariance changes over time. The large downside to this technique is the amount of computational time it takes to complete. Another potential technique is to use Gaussian Process regression. The light curve containing only scintillation noise can be written as:

\begin{equation}
    \label{equ:lightcurve_distribution_scintonly}
    \textbf{y}_{s} \sim \mathcal{N}\left(\textbf{T}, \mathbf{\Sigma}_{s}\right)
\end{equation}

\noindent where $\mathbf{\Sigma}_{s}$ is the covariance due to scintillation only. Since realizations of this distribution are perfectly correlated with the scintillation observed light curve, $y_{s+p}$, the covariance between these two distributions is $\mathbf{\Sigma}_{s}$. Now, these two distributions can be expressed as a joint Gaussian distribution

\begin{align}
\label{equation:jointlikelihood}
\begin{pmatrix}\textbf{y}_{s+p}\\
\textbf{y}_{s}
\end{pmatrix} &\sim  \mathcal{N}
\begin{bmatrix}
\begin{pmatrix}
\textbf{T}\\
\textbf{T}\\
\end{pmatrix}\, ,&
\begin{pmatrix}
\mathbf{\Sigma}_{s+p} & \mathbf{\Sigma}_{s} \\
\mathbf{\Sigma}_{s} & \mathbf{\Sigma}_{s}\\
\end{pmatrix}
\end{bmatrix}
\end{align}
\\

\noindent Predictions of the scintillation noise given the total noise can be made using the posterior
%fix
\begin{equation}
    \label{equ:scintillation_posterior1}
    p(\textbf{y}_{s}|\textbf{y}_{s+p},\textbf{T}) = \mathcal{N}(\textbf{T}, \mathbf{\Sigma}_{s|s+p})
\end{equation}
\begin{equation}
	\label{equ:scintillation_posterior2}
	r_{s|s+p} = \mathbf{\Sigma}_{s} \mathbf{\Sigma}_{s+p}^{-1} (\textbf{y}_{s+p} - \textbf{T})
\end{equation}
\begin{equation}
	\label{equ:scintillation_posterior3}
    \mathbf{\Sigma}_{s|s+p} = \mathbf{\Sigma}_s - \mathbf{\Sigma}_{s} \mathbf{\Sigma}_{s+p}^{-1} \mathbf{\Sigma}_{s}
\end{equation}

A future paper will explore the best way to correct for this scintillation bias. 
\\

As mentioned in the assumptions of Section \ref{ssec: simulating_ground_based_tspec}, time-correlated noise is not simulated. The obvious problem with time-correlated noise is that if it is left untreated, it will induce a bias in the inferred astrophysical signal. If the time-correlated noise and its wavelength dependence can be modeled as precisely as the scintillation noise, then nothing about the scintillation estimation scheme should change. Many authors have had success removing time-correlated noise from exoplanet transit light curves \citep{Gibson2012, Diamond-Lowe_2023, Ahrer2022, Panwar2022, McGruder_2022}. However, it is not clear that these techniques can remove the time-correlated noise without perturbing the transit signal\footnote{Note that differential spectrophotometry, a popular method for isolating and removing common-mode noise induced by the atmosphere, will not affect scintillation estimation. Although both stars will see different scintillation signals, when they are combined their covariances are summed.}, In addition, there are subtleties that arise when defining the transpose and inverse of a 4-dimensional matrix. Since no time-correlated noise is present in this simulation, calculating the inverse reduces to calculating the inverse for $N$ $M \times M$ matrices. In this work, the inverse is defined such that $AA^{-1} := I_{t} \otimes I_{\lambda}$, where $I_{\lambda}$ and $I_{t}$ are $2$-dimensional identity matrices with sizes corresponding to the number of wavelengths and the number of exposures. The transpose can be defined as $[\textbf{A}]^{T}_{ijkl} = [\textbf{A}]^{T}_{jilk}$ where $ij$ represent the either the time or wavelength indices and $kl$ represent the other. Due to the complexities of modeling both temporal and chromatic correlations in spectrophotometry, it is best to limit any sources of noise to below the level of the Poisson noise or to ones that are accurately described by simple models.

\subsubsection{A note on photometry}
\label{ssec:photometry}

Knowledge of the scintillation covariance can also be used to reduce scintillation noise in photometry. In the simplest case, a two channel photometer where one channel is used for scintillation correction is an option. Although this method would certainly increase the precision, biases from the scintillation noise may remain due to only having one reference data point. Another interesting way to reduce scintillation noise comes when using small (< 0.5 meter) telescopes. Since the covariance decreases exponentially with telescope diameter, if an object's flux is recorded through widely separated filters such that the scintillation patterns are uncorrelated in each filtered bandpass, the scintillation noise is decreased by a factor of $\sqrt{N_{\text{filters}}}$. This will depend on the wind-direction, wind speed, and airmass and therefore, for some fraction of nights this can be a particularly powerful technique to reduce the effects of scintillation. When combining many small telescopes that employ this approach, \citep{Bakos2004,Wheatley2018} very high precision photometry may be achieved under common observing conditions.

\section{Conclusions}

The dominant source of noise for ground-based observations of bright stars is often scintillation due to high-altitude turbulence in the atmosphere. This work presents a technique to improve the accuracy and precision of ground-based spectrophotometry by isolating and removing the effects of scintillation noise. By observing light from bright stars at multiple wavelengths,  \textit{nearly} achromatic scintillation signal is produced. In addition to reducing the precision of an observation, it also induces a bias in wavelength space due to it being achromatic. By developing an analytical formulation for the covariance at two wavelengths, accurate and precise measurements of astrophysical signals can be made for bright objects. This technique is best applied for observations that are dominated by scintillation and Poisson noise. The results of this simulation point to this technique being able to correct for the effects of scintillation below the level of Poisson noise when the total noise is dominated by scintillation.
\\

The primary advantage of isolating scintillation noise in this way is that it is completely data-driven - it only depends on the observed data (i.e. light curves, time, wavelength, and airmass). Although this method fundamentally relies on the fact that the dominant noise sources in the light curves are caused by scintillation and Poisson noise, additional noise sources should pose no issue as long as their temporal and wavelength covariance is well understood or can be well approximated. If observations from an instrument produce light curves with any time-correlated noise or additional wavelength-correlated noise, updates to the covariance are needed since original fit to each light curve will likely be even less accurate. Therefore, instruments that suppress or eliminate the effects of time-correlated noise may be particularly suited for this  technique.
\\

The accuracy limits of this technique will be tested on-sky during the commissioning phase of Henrietta \citep{Williams2022}, a new low-resolution spectrograph for the 1-meter Swope Telescope at Las Campanas observatory. Henrietta is specifically designed to take advantage of this scintillation correction technique by reducing all other sources of noise so that scintillation and Poisson noise are the dominant sources of error. With this scintillation correction procedure, Henrietta's goal is to demonstrate routine, Poisson-noise limited transmission spectroscopy of exoplanets around bright stars. Henrietta is projected to be on sky in December 2025.

\section*{Acknowledgements}

The author thanks Nicholas Kondiaris for his guidance and helpful comments. The author also thanks James Osborn for his helpful comments along with providing access to the Stero-SCIDAR data used for the simulation. This material is based on substantial funding from Carnegie Science as a graduate student and postdoctoral research fellow. This research was funded by the Heising-Simons Foundation through grant 2021-2614. This material is based upon work supported by the National Science Foundation under Grant No. 2206374. Author recognizes generous support from the Ahmanson Foundation and the Mt. Cuba Foundation. 

%%%%%%%%%%%%%%%%%%%%%%%%%%%%%%%%%%%%%%%%%%%%%%%%%%
\section*{Data Availability}

The code used in this simulation can be found in the Github repository: https://github.com/williamscommajason/Scintillation.

\newpage



\bibliographystyle{mnras}
\bibliography{thesisbib3} % if your bibtex file is called example.bib

%\begin{thebibliography}{99}
%\end{thebibliography}

%%%%%%%%%%%%%%%%%%%%%%%%%%%%%%%%%%%%%%%%%%%%%%%%%%

%%%%%%%%%%%%%%%%% APPENDICES %%%%%%%%%%%%%%%%%%%%%
\newpage

\appendix
\section{Derivation of scintillation equations}
\label{sec:scintillationderivationappendix}
\subsection{Short exposure scintillation equations}
\label{sec:scintillationvariance}
Following \citep{Roddier1981} and \citep{Tokovinin2002}, the power of intensity fluctuations due to scintillation is given by 

\begin{equation}
    \label{equation:scintillationvariance}
    \mathbf{\sigma} = \int_0^{\infty} dz \, C_n^{2}(z) W(\lambda, z)
\end{equation}

\noindent where $W(\lambda,z)$ is the altitude weighting function

\begin{equation}
    \label{equation:scintillationvariancepowerspectrum}
    W(\lambda,z) = \frac{9.62}{\lambda^2} \int_0^{\infty} df \, f\, \Phi(f) \sin^2(\pi \lambda z f^2) \, \left(\frac{2 J_{1}(\pi D f)}{(\pi D f)}\right)^2
\end{equation}

\noindent and $\lambda$ is the wavelength of light, $f$ is the spatial frequency of refractive index fluctuations, $z$ is the altitude, $D$ is the receiving aperture diameter, and $C_{n}^{2}(z)$ is the vertical distribution of refractive index fluctuations (referred to as the turbulence profile). $\Phi(f)$ is power spectrum of refractive index fluctuations, assumed to be a Kolmogorov spectrum equal to $f^{-11/3}$. Equation \ref{equation:scintillationvariance} has been approximated several times \citep{Roddier1981, Kenyon2006, Osborn2015}, and this works introduces another approximation that will serve to make future computations easier.

The quantity $\left(\frac{2 J_{1}(\pi D f)}{(\pi D f)}\right)^2$ is known as the aperture averaging factor (or aperture filter), which accounts for the averaging over the intensity fluctuations over a receiving aperture $D$. This factor can be approximated by an exponential

\begin{equation}
    \label{equation:lowfrequencyapproximation}
    \left(\frac{2 J_{1}(\pi D f)}{(\pi D f)}\right)^2 \approx e^{-b^2 D^2 f^2}
\end{equation}

\noindent where $b \sim 2.7$ minimizes the difference in integrated area between both functions. Making the substitution $x = f^2$ and using the equations\footnote{Note that given $\mu = -5/6$, these equations do not converge individually. However, the difference of these two equations converges since $\lim_{x \to 0} \frac{1 - \cos(x)}{x^{11/6}} \approx x^{1/6}$}

\begin{equation}
    \label{equation:sinsquaredtocosine}
    \sin^2(x) = \frac{1}{2}\left(1 - \cos(2x)\right)
\end{equation}

\begin{equation}
\int_0^\infty x^{\mu - 1}\, e^{-\beta x} \, \cos(\zeta \,x)\,dx = \frac{\Gamma(\mu)}{(\zeta^2 + \beta^2)^{\mu\,/2}} \cos(\mu \tan^{-1}(\zeta/\beta))
\label{equation:gradshteyn1}
\end{equation}

\begin{equation}
\int_0^\infty  x^{\mu - 1}\, e^{-\beta x} \,dx = \frac{\Gamma(\mu)}{\beta^\mu}
\label{equation:gradshteyn2}
\end{equation}
\\

\noindent and the fact that $\lambda z << 1$, $W(f)$ reduces to 

\begin{equation}
    \label{equation:scintillationvarianceapproximation}
    W(\lambda , z) \sim 13.83 \, D^{-7/3} \sec(\xi)^3 z^2
\end{equation}

\noindent Apart from the multiplicative constant (which can be changed by changing $b$), this approximation leads to the exact functional behavior ($D^{-7/3}, \sec(\xi)^3, z^2$ and no dependence on $\lambda$) derived by previous authors \citep{Roddier1981,Kenyon2006,Osborn2015}. However, the primary point in introducing this approximation is to show that it leads to similar functional behavior while producing an analytical solution.

\subsubsection{Scintillation covariance for non-dispersive path lengths}
\label{ssec:scintillationcovariance}
Following the definitions from \citep{Ryan1998, Kornilov2011a}, the covariance of scintillation is 

\begin{equation}
    \label{equation:scintillationcovariance}
    \mathbf{\sigma}_{\lambda_{1},\lambda_{2}}^{2} = \int_0^{\infty} dz \, C_n^{2}(z) \, X(\lambda_1, \lambda_2, z)
\end{equation}
\\
where $X(\lambda_1, \lambda_2, z)$ is the covariant weighting function defined as 

\begin{align}
    \label{equation:scintillationcovariancepowerspectrum}
    X(\lambda_1,\lambda_2,z, D) = \frac{9.62\, \sec(\xi)}{\lambda_1  \lambda_2} \int_0^{\infty} df \, f \Phi(f) \sin(\pi \lambda_1 z f^2) & \nonumber \\
    \qquad \times \, \sin(\pi \lambda_2 z f^2) \left(\frac{2 J_{1}(\pi D f)}{(\pi D f)}\right)^2
\end{align}

\noindent The integral can be reparameterized in terms of one wavelength and the wavelength ratio $r = \frac{\lambda_1}{\lambda_2}$, using the cosine addition/subtraction formulas,

\begin{align}
    \label{equation:scintillationcovariancepowerspectrum2}
    X(\lambda_1,\lambda_2,z, D) = \frac{9.62\, \sec(\xi)}{2\,\lambda_1  \lambda_2} \int_0^{\infty} df \, f \Phi(f) \left(\frac{2 J_{1}(\pi D f)}{(\pi D f)}\right)^2 & \nonumber \\
    \qquad \times \, \left[\cos(\pi \lambda_2 (1 - r) z_{\xi} f^2) - \cos(\pi \lambda_2 (1 + r) z_{\xi} f^2)\right]
\end{align}

\noindent Applying the aperture filter approximation in Equation \ref{equation:lowfrequencyapproximation}, Equation \ref{equation:scintillationcovariancepowerspectrum2} can be solved exactly using Equation \ref{equation:gradshteyn2} with $\cos(x)$ written in its exponential form

\begin{align}
    \label{equation:scintillationcovariancepowerspectrum3}
    X(\lambda_1,\lambda_2,z, D) \sim \frac{9.62 \, \sec(\xi)}{2\,\lambda_1  \lambda_2} \, \Re\Biggl[ \int_0^{\infty} df\, f^{-8/3} & \nonumber \\
    \qquad \times \, \exp\left[-f^{2}(b^2 D^2 - i\pi \lambda_2 (1 - r) z_{\xi})\right] \nonumber \\
    -\, \exp\left[-f^{2}(b^2 D^2 - i\pi \lambda_2 (1 + r) z_{\xi} )\right] \Biggr]
\end{align}

\begin{multline}
    \label{equation:scintillationcovariancepowerspectrum4}
    X(\lambda_1,\lambda_2,z, D) \sim \frac{9.62 \, \Gamma(-\frac{5}{6}) \, \sec(\xi)}{4\,\lambda_1  \lambda_2}  \\
     \times \, \Re \left[\left(b^2 D^2 - i\pi \lambda_2 (1 - r) z_{\xi}\right)^{5/6} - \left(b^2 D^2 - i\pi \lambda_2 (1 + r) z_{\xi}\right)^{5/6} \right]
\end{multline}

\subsubsection{Scintillation covariance for dispersive paths}

As mentioned in Section \ref{ssec:groundbasedscintillationcovariance}, the dispersion of the atmosphere also influences the covariance of scintillation on ground-based telescopes. For short exposures, the expression is

\begin{multline}
    \label{equation:scintillation_dispersion_covariance_shortexp2}
    \sigma_{\lambda_{1}, \lambda_{2}}^{2}(\rho) \approx  \frac{0.32\, \sec(\xi)}{2\lambda_1  \lambda_2} \int_0^{\infty} df \, f^{-8/3} \int^\infty_0 dz\,C_n^2(z) \\[7pt] \times \left[\cos\left(\pi \lambda_2(1 - \frac{\lambda_1}{\lambda_2})\, z\, \text{sec}(\xi)\,f^2\right) - \cos\left(\pi \lambda_2(1 + \frac{\lambda_1}{\lambda_2})\, z\, \text{sec}(\xi)\, f^2\right)\right] \\[7pt] \times J_0\left(2\pi f \rho(\lambda_1, \lambda_2, \xi, z) \right)  
  e^{-b^2 D^2 f^2}
\end{multline}

\noindent where $r = \frac{\lambda_1}{\lambda_2}$. With the approximation for the aperture filter, Equation \ref{equation:scintillation_dispersion_covariance_shortexp2} can be analytically integrated via the integral representation of the Laguerre function \citep{Abramowitz1965}

\begin{equation}
\label{equation:laguerrefunction}
L_n^{(\alpha)}(x) = \frac{e^x x^{-\alpha/2}}{n!}\int_0^\infty \, dt \, e^{-t} \, t^{n + \alpha/2} \, J_{\alpha}\left(2\sqrt{x t}\right)
\end{equation}

\noindent for $\alpha \geq -1$. Writing cosine in its exponential form, making the substitutions $$t = f^2 \left((bD)^2 - 2 \pi i\,z_{\xi} \lambda_2 (1\pm \frac{\lambda_1}{\lambda_2})\right)$$ $$x^\pm =  \frac{(\rho(\lambda_1, \lambda_2, \xi, z))^2}{(bD)^2 - 2\pi i\,z_{\xi} \lambda_2 (1\pm \frac{\lambda_1}{\lambda_2}) }$$ 
\\
and using the identity $$ L_{-n}(x) = e^{x}L_{n-1}(-x) $$ results in 

\begin{multline}
    \label{equation:scintillation_dispersion_covariance_shortexp3}
     \sigma_{\lambda_{1}, \lambda_{2}}^{2}(\rho) \approx \frac{0.32\, \Gamma(-5/6)}{4\lambda_1  \lambda_2} \int^\infty_0 dz_{\xi}\, C_n^2(z_{\xi}) \\[7pt] \times \mathfrak{Re} \, \Big[G_{-}(z_{\xi}\,{;}\, x^+) - G_{+}(z_{\xi}\,{;}\, x^{-}) \Big]
\end{multline}

\noindent where 

\begin{equation}
G_{\pm}(z_{\xi}\,{;}\, x^{\pm}) = L_{5/6}(-x^{\pm}) \left((bD)^2 - 2\pi i\,z_{\xi} \lambda_2 (1 \pm \frac{\lambda_1}{\lambda_2}) \right)^{5/6}
\end{equation}

\subsection{Long exposure scintillation equations}

The previous subsections outlined expressions for the variance and covariance in the short-exposure regime, realized in \citep{Kornilov2011c} as $w\tau << D$ where $w$ is the projected wind-speed across the turbulent layer and $\tau$ is the exposure time.\footnote{The short-exposure regime is essentially a restatement of Taylor's frozen turbulence hypothesis \citep{Taylor1938}.} For the long exposures typical in ground-based spectrophotometry, the turbulence profile is averaged across the line of sight by the wind speed, leading to an averaging down of the observed intensity fluctuations. Since atmospheric turbulence is assumed to be isotropic and homogeneous \citep{Kolmogorov1941, Roddier1981}, the wind direction does not introduce a preferred direction. Following the formulation from \citep{Tokovinin2002}, the effect of exposure time on the variance and covariance amounts to multiplying the weighting functions by the wind smoothing filter, $A_{w}(f)$:

\begin{equation}
    \label{equation:sincsquared}
    A_{w}(f;w,\tau) = \int_0^{2\pi} d\phi \, \text{sinc}^2(w\tau f \cos(\phi)) 
\end{equation}

\noindent When $w\tau f >> 1$, $A_{w}(f;w,\tau)$ tends to $1/\pi w \tau f$ (\citep{Tokovinin2002}). This condition is easily met for most conventional spectrophotometry where exposure times are long (i.e. $w\tau >> D).$. Adding this into the expressions for the scintillation variance and covariance weighting functions, Equation \ref{equation:scintillationvariancepowerspectrum} becomes 

\begin{equation}
    \label{equation:scintillationvariancespectrum_longexp}
    W(\lambda,z) = \frac{9.62}{\pi w(z) \tau \lambda^2} \int_0^{\infty} df \, f^{-8/3} \sin^2(\pi \lambda z f^2) \, \left(\frac{2 J_{1}(\pi D f)}{(\pi D f)}\right)^2
\end{equation}

\noindent Equation \ref{equation:scintillationcovariancepowerspectrum2}  becomes 

\begin{align}
    \label{equation:scintillationcovariancepowerspectrum_longexp}
    X(\lambda_1,\lambda_2,z) = \frac{9.62\, \sec(\xi)}{2 \pi w(z) \tau \,\lambda_1  \lambda_2} \int_0^{\infty} df \, f^{-8/3} \left(\frac{2 J_{1}(\pi D f)}{(\pi D f)}\right)^2 & \nonumber \\
    \qquad \times \, \cos(\pi \lambda_2 (1 - r) z_{\xi} f^2) - \cos(\pi \lambda_2 (1 + r) z_{\xi} f^2)
\end{align}

\noindent Making the same approximations and using the same integrals in previous sections, Equation \ref{equation:scintillationvariancespectrum_longexp} becomes

\begin{equation}
    \label{equation:scintillationvariancespectrum_longexp1}
    W(\lambda, z) \sim 10.58 D^{-4/3} \sec(\xi)^3 \frac{z^2}{w(z)\,\tau}
\end{equation}

\noindent and Equation 
\ref{equation:scintillationcovariancepowerspectrum_longexp} becomes

\begin{multline}
    \label{equation:scintillationcovariancespectrum_longexp1}
    X(\lambda_1,\lambda_2,z) \sim \frac{9.62 \, \Gamma(-\frac{4}{3}) \, \sec(\xi)}{4 \pi w(z) \tau \,\lambda_1  \lambda_2}  \\
     \times \, \mathfrak{Re} \left[\left(b^2 D^2 - i\pi \lambda_2 (1 - r) z_{\xi}\right)^{4/3} - \left(b^2 D^2 - i\pi \lambda_2 (1 + r) z_{\xi}\right)^{4/3} \right]
\end{multline}

\noindent which are the same functional dependencies posited in \citep{Kornilov2011a} and \citep{Kornilov2011b}. The effect of longer exposures is to decrease the scintillation (co)variance. In this long-exposure regime, scintillation is essentially white noise in the temporal domain as observed and postulated by many authors \citep{Roddier1981,Dravins1998,Kenyon2006,Kornilov2011c,Osborn2015,Fohring2019}. 
\\
\subsubsection{Scintillation covariance for dispersive paths}

\noindent Starting from Equation \ref{equation:sincsquared1}, using the Binomial Theorem and a cosine difference formula gives 

\begin{multline}
    \label{equation:windaveragingfilter_parallel1}
    A_{w}(f, \rho ,w,\tau, \theta) = \sum_{n = 0}^{\infty} \frac{(2\pi i f \rho)^n}{2\pi\,n!} \sum_{k=0}^{n} \binom{n}{k} \, \cos^{n-k}(\theta) \sin^{k}(\theta) \\[7pt] \int_{0}^{2\pi} d\phi \, \cos^{n-k}(\phi)\sin^{k}(\phi) \, \text{sinc}^2(w\tau f \cos(\phi)) 
\end{multline}
\\

\noindent The integral over $\phi$ can be rewritten as 
\begin{multline*}
\int_{0}^{2\pi} d\phi \, \cos^{n-k}(\phi)\sin^{k}(\phi) \, \text{sinc}^2(w\tau f \cos(\phi)) = \\[7pt]
F(n,k) \, \int_{3\pi/2}^{2\pi} d\phi \, \cos^{n-k}(\phi) \, \sin^{k}(\phi) \, \text{sinc}^{2}(w \tau f \cos(\phi))
\end{multline*}
\\

\noindent where $F(n,k) = (1/2\pi)((-1)^{n-k} + 1) ((-1)^k + 1)$. The key to this integral is to rewrite the $\text{sinc}^2$ term as 

\begin{equation}
    \text{sinc}^2(...) = \frac{1}{(w\tau)^2 f \cos(\phi)}\int_0^{w\tau} dt \, \sin(2 f t \cos(\phi))
\end{equation}

\noindent so that Equation \ref{equation:windaveragingfilter_parallel1} becomes 
\begin{multline}
    \label{equation:windaveragingfilter_parallel2}
    A_{w}(f,w,\tau) = 
    \sum_{n = 0}^{\infty} \frac{(2\pi i \rho)^n f^{n-1}}{(w^2 \, \tau^2)n!} \sum_{k=0}^{n} G(n,k,\theta) \\[7pt] \int_{0}^{w\tau} dt \int_{3\pi/2}^{2\pi} d\phi \, \cos^{n-k-1}(\phi) \sin^{k}(\phi) \sin(2 f t \cos(\phi))
\end{multline}
\\

\noindent where $G(n,k,\theta) = \binom{n}{k} \, \cos^{n-k}(\theta) \sin^{k}(\theta) F(n,k)$. \noindent Making the substitution $y = \cos(\phi)$ it becomes

\begin{multline}
    \label{equation:windaveragingfilter_parallel3}
    A_{w}(f,w,\tau) = -\sum_{n = 0}^{\infty} \frac{(2\pi i \rho)^n f^{n-1}}{(w^2 \, \tau^2)n!} \sum_{k=0}^{n} G(n,k,\theta)  \\[7pt] \int_{0}^{w\tau} dt  \int_{0}^{1} dy \, \frac{y^{n-1} \, \sin(2 f t y)}{\sqrt{(1-y^{2})^{k-1}}}
\end{multline}

\noindent Using an integral from Page 442 of \cite{Gradshteyn2015}, the $y$ integral can be solved to give

\begin{multline}
    \label{equation:windaveragingfilter_parallel4}
    A_{w}(f,w,\tau) = -\sum_{n = 0}^{\infty} \frac{(2\pi i \rho)^n f^{n}}{(w^2 \, \tau^2)n!} \sum_{k=0}^{n} G(n,k,\theta) \, \frac{\Gamma(\frac{k+1}{2})\,\Gamma(\frac{n-k+1}{2})}{\Gamma(\frac{n + 2}{2})} \\[7pt]
    \times \int_{0}^{w\tau} dt \, t \ {}_{1}F_{2} \left[\left\{\frac{n-k+1}{2}\right\},\left\{\frac{3}{2}, \frac{n+2}{2}\right\},{-f^2 v^2 \tau^2}\right]
\end{multline}
\\

\noindent where ${}_{1}F_{2}$ is the generalized hypergeometric function defined on Page 1010 of \citep{Gradshteyn2015}. The integral over $t$ is completed using Equation $7.522.1$ in \cite{Gradshteyn2015} and amounts to 

\begin{multline}
    \label{equation:windaveragingfilter_parallel5}
    A_{w}(f,w,\tau) = -\sum_{n = 0}^{\infty} \frac{(2\pi i \rho)^n f^{n}}{n!} \sum_{k=0}^{n} G(n,k,\theta) \, \frac{\Gamma(\frac{k+1}{2})\,\Gamma(\frac{n-k+1}{2})}{\Gamma(\frac{n + 2}{2})}  \\[7pt]
    \times {}_{2}F_{3} \left[ \left\{1,\frac{n-k+1}{2}\right\}, \left\{2,\frac{3}{2}, \frac{n+2}{2}\right\},{-f^2 v^2 \tau^2}\right]
\end{multline}

Since the upper and lower parameters of ${}_{2}F_{3}$ are separated by an integer, a theorem from \cite{Withers2014} reformulates this as 
\begin{multline}
    \label{equation:windaveragingfilter_parallel6}
    A_{w}(f,w,\tau) = -\sum_{n = 0}^{\infty} \frac{(2\pi i \rho)^n f^{n-2}}{(2\, w^2 \, \tau^2)\, n!} \sum_{k=0}^{n} G(n,k,\theta) \, \frac{\Gamma(\frac{k+1}{2})\,\Gamma(\frac{n-k-1}{2})}{\Gamma(\frac{n}{2})}  \\[7pt]
    \times {}_{1}F_{2}\left[\left\{\frac{n-k-1}{2}\right\},\left\{\frac{1}{2}, \frac{n}{2}\right\},{-f^2 v^2 \tau^2} - 1\right]
\end{multline}

This expression for the wind-filter can now be substituted into the expression for the covariance to give
\begin{multline}
    \sigma_{\lambda_1, \lambda_2}^{2}(\rho, w, \tau, \theta) = \sum_{n = 0}^{\infty} \frac{(2\pi i \rho)^n}{(w^2 \tau^2 n!)} \sum_{k=0}^{n} G(n,k,\theta) \, \frac{\Gamma(\frac{k+1}{2})\,\Gamma(\frac{n-k-1}{2})}{\Gamma(\frac{n}{2})} \\[7pt] \qquad \frac{0.32}{4\pi\lambda_1  \lambda_2} \int_{0}^{H/\sec(\xi)} dz_\xi \, C_{n}^{2}(z_{\xi}) \int_0^{\infty} df  \, f^{n -14/3} \sin(\pi \lambda_1 z_{\xi} f^2)  \\[7pt]
    \times \sin(\pi \lambda_2 z_{\xi} f^2)\, {}_{1}F_{2}\left[\left\{\frac{n-k-1}{2}\right\}, \left\{\frac{1}{2}, \frac{n}{2}\right\},{-f^2 v^2 \tau^2} - 1\right]  \left(\frac{2 J_{1}(\pi D f)}{(\pi D f)}\right)^2
\end{multline}

The value of the wind-filter at $n = 0$ is the expression for the long exposure variance and is approximated well by Equation \ref{equation:scintillationvariancespectrum_longexp1} \footnote{For $n = 0$ to be defined, the ${}_{2}F_{3}$ in Equation \ref{equation:windaveragingfilter_parallel5} representation must be used.}. Making the aperture averaging approximation and using Equation 7.522.9 in \cite{Gradshteyn2015}, the result is 

\begin{multline}
\label{equation:final_long_exposure_covariance}
    \sigma_{\lambda_1, \lambda_2}^{2}(\rho, w, \tau, \theta) \approx \frac{0.32}{16\pi\lambda_1  \lambda_2} \int_{0}^{H/\sec(\xi)} dz_\xi \, C_{n}^{2}(z_{\xi}) \\[7pt] \qquad \sum_{n = 0}^{\infty} \frac{(2\pi i \rho)^n}{(w^2 \tau^2 n!)} \sum_{k=0}^{n} G(n,k,\theta) \frac{\Gamma(\frac{k+1}{2})\,\Gamma(\frac{n-k-1}{2}) \, \Gamma(\frac{n}{2} - \frac{11}{6})}{\Gamma(\frac{n}{2})} \\[7pt] 
    \times \mathfrak{Re} \left[F_{n}^{-}(\tau, w, z_{\xi}) - F_{n}^{+}(\tau, w, z_{\xi}) \right]
\end{multline}

\noindent where 
\begin{multline}
\label{equation:hypergeom}
F_{n}^{\pm}(\tau, w, z_{\xi}) = \left(b^2 D^2 - i\pi \lambda_2 \left(1 \pm \frac{\lambda_1}{\lambda_2}\right) z_{\xi}\right)^{11/6 - \frac{n}{2}} \\[7pt] {}_{1}F_{2} \left[\left\{\frac{n}{2} - \frac{11}{6}, \frac{n-k-1}{2}\right\}, \left\{\frac{1}{2}, \frac{n}{2}\right\},{\frac{-w^2 \tau^2}{b^2 D^2 - i \pi \lambda_{2} (1 \pm \frac{\lambda_1}{\lambda_2}) z_{\xi}}}-1\right]
\end{multline}
\\

\noindent and 

\begin{multline}
\label{equation:longexposure_covariantweightingfunction_final}
Y(z, \lambda_1, \lambda_2, \tau, w, \theta) = \frac{0.32 \, \sec(\xi) \, \mathfrak{Re} \left[F_{n}^{-}(\tau, w, z_{\xi}) - F_{n}^{+}(\tau, w, z_{\xi}) \right]}{16\pi\lambda_1  \lambda_2}  \\[7pt] \times \sum_{n = 0}^{\infty} \frac{(2\pi i \rho)^n}{(w^2 \tau^2 n!)} \sum_{k=0}^{n} G(n,k,\theta) \frac{\Gamma(\frac{k+1}{2})\,\Gamma(\frac{n-k-1}{2}) \, \Gamma(\frac{n}{2} - \frac{11}{6})}{\Gamma(\frac{n}{2})} 
\end{multline}
\\

%Remember to FIX zenith dependence of velocity!!
%Fix minus signs
%fix wolfram referencesAZQ
%fix figure 3 shows correlation not weighting function
%talk about diagonal of covariance gives the uncertainties talk about noise in wavelength space and time space.
Evaluation of the hypergeometric function ${}_{2} F_{2}$ in Equation \ref{equation:hypergeom} can be quite slow, especially when it needs to be called multiple times (in the case of multiple wavelengths, in an MCMC sampler, etc.). Fortunately, \citep{Olver2010} provides an expression for the expansions for a large variable when $|\text{ph}(-z)| \leq \pi$

\begin{multline}
    {}_{q} F_{q} \left(\begin{matrix}a_1,...,a_q \\b_1,...,b_q \end{matrix};-z\right) \approx \left(\prod_{l = 1}^{q} \Gamma(b_l) \middle/ \prod_{l = 1}^{q} \Gamma(a_l)\right) H_{q,q}(z) 
\end{multline}

where 
\begin{multline}
H_{q,q}\left(\begin{matrix}a_1,...,a_q \\b_1,...,b_q \end{matrix};z\right)= \sum_{m = 1}^{p} \sum_{k = 0}^\infty \frac{(-1)^k}{k!} \Gamma(a_{m} + k) \, z^{-a_{m} - k} \\[7pt] \left(\prod_{\substack{l = 1 \\ l \neq m}}^p \Gamma(a_l - a_m - k) \middle/ \prod_{l = 1}^q \Gamma(b_l - a_m - k) \right)
\end{multline}

For $|\mathfrak{Re}(z)| \sim 10 $, this formulation is accurate to within $5-15\%$ depending on the parameters, with the accuracy increasing exponentially as z grows. Since $\frac{v^2 \tau^2}{b^2 D^2}$ will typically be much greater than 10, this approximation will be used to dramatically improve the speed of evaluation ($ > 25\times$ faster).

\section{Simulating Ground-based Transit Spectroscopy}
\label{ssec: simulating_ground_based_tspec}

\begin{figure}
 \includegraphics[width=\columnwidth]{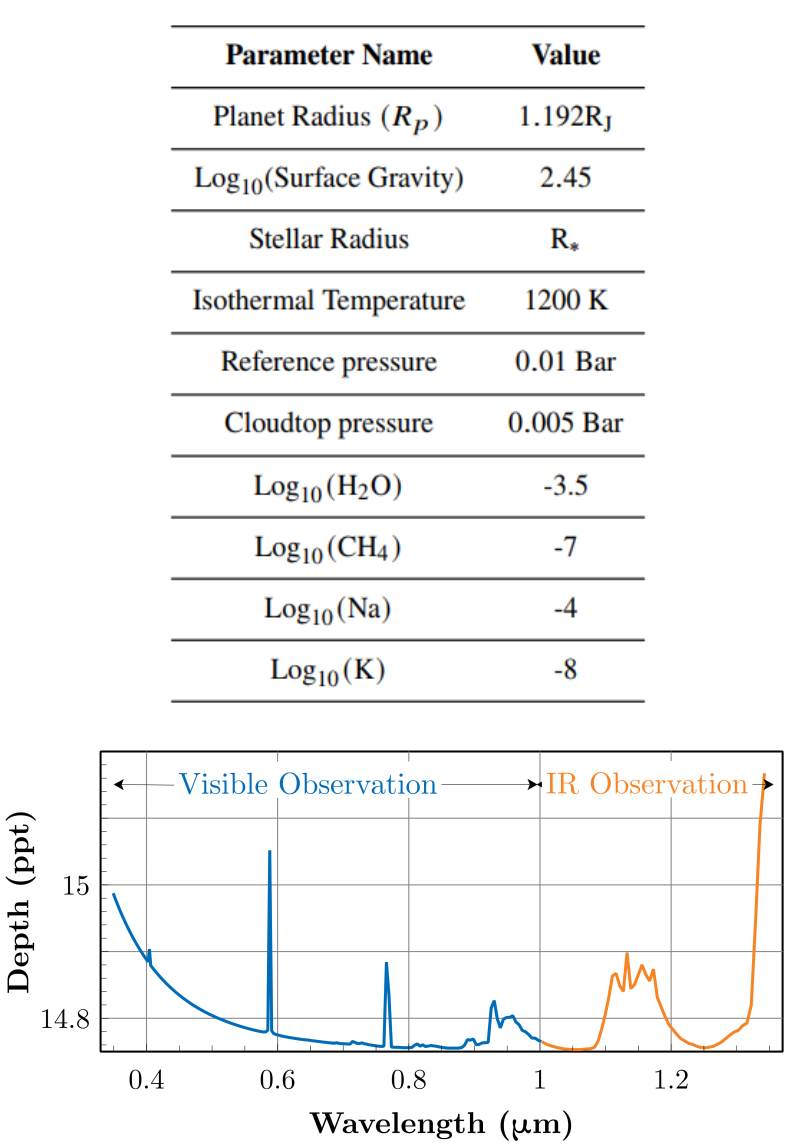}
 \caption{\textbf{Planetary and Atmospheric parameters along with resulting transmission spectrum from petitRADTRANS.} The regions from $0.35 {\mu}m$ to $1 {\mu}m$ and $1 {\mu}m$ to $1.35 {\mu}m$ are 'observed' separately to assess how scintillation can affect two non simultaneous observations.}
 \label{fig:atmospheric_parameters_and_spectrum}
\end{figure}

Exoplanet transit spectroscopy - one of the most successful techniques to characterize an exoplanet's atmosphere - requires observations with sufficient spectral resolutions (typically $\mathcal{R} > 100$, \citep{Tinetti2013}) at high precisions (1- 100 ppm per transit duration). The observational challenges imposed by transit spectroscopy thus serve as a useful benchmark to develop a scintillation suppression scheme since a) measuring a transit requires precisions on the order of hundreds to thousands of parts-per-million and b) the changes of the transit depth across wavelength are on the order of tens of hundreds of parts per million, similar to the effects of scintillation. To generate a transit spectrum, a model atmosphere, wavelength range, and spectral resolution must be chosen. Using the exoplanet atmosphere radiative transfer code petitRADTRANS \citep{Molliere_2019}, an isothermal exoplanet atmosphere was simulated. Each light curve is generated using the transit light curve modeling package \textsc{batman} \citep{Kreidberg2015a}. The transit model for each wavelength $m$ is represented as
\begin{equation} 
\label{equation:Transitmodel}
\textbf{T}_{m} = F\left((\overrightarrow{\RpRs}), i, a/R_{*}\right)
\end{equation}
\noindent The planetary parameters and the resulting transmission spectrum is plotted in Figure \ref{fig:atmospheric_parameters_and_spectrum}.
\\

%\begin{table}
%\label{table:atmospheric_parameters}
%\caption{\textbf{Atmospheric and planetary parameters %for the simulated transmission spectrum.}}
%\centering
%\begin{tabular}{|c|c|}
%\hline
%\textbf{Parameter Name}               & \textbf{Value} %\\ \hline
%Planet Radius $(R_{p})$     & $1.192 \text{R}_{\text{J}}$ \\ \hline
%$\text{Log}_{10}$(Surface Gravity) & $2.45$           %\\ \hline
%Stellar Radius              & $\text{R}_*$           \%\ \hline
%Isothermal Temperature       & $1200$ K         \\ \hline
%Reference pressure           & $0.01$ Bar       \\ \hline
%Cloudtop pressure            & $0.005$ Bar       \\ \hline
%$\text{Log}_{10}(\text{H}_{2}\text{O})$                  & -3.5           \\ \hline
%$\text{Log}_{10}(\text{CH}_{4})$                       %& -7             \\ \hline
%$\text{Log}_{10}(\text{Na})$                       & -4             \\ \hline
%$\text{Log}_{10}(\text{K})$                       & -8             \\ \hline
%\end{tabular}
%\end{table}

The noise will be simulated in two observations that span different wavelength ranges - one from $0.35 {\mu}m$ to $1 {\mu}m$ (called the 'visible' observation) and the other from $1 {\mu}m$ to $1.35 {\mu}m$ (called the 'infrared' observation). This is meant to mimic a realistic scenario in transmission spectroscopy where observations are combined from different instruments on the same night, different observing modes on different nights, etc.
\\

To simulate time-varying scintillation noise, the vertical optical turbulence profile $C_n^2(z)$, the vertical wind profile $w(z)$, and the wind-direction $\theta(z)$ needs to be calculated at every time-step. The most comprehensive dataset that contains all of these parameters is the ESO/Durham University Stereo-SCIDAR dataset gathered at Paranal Observatory \citep{Osborn2018}. Observing for a total of 83 nights across 22 months, the dataset produced real-time observations that characterized the optical turbulence above Paranal Observatory. 
\\

\begin{figure}
 \includegraphics[width=\columnwidth]{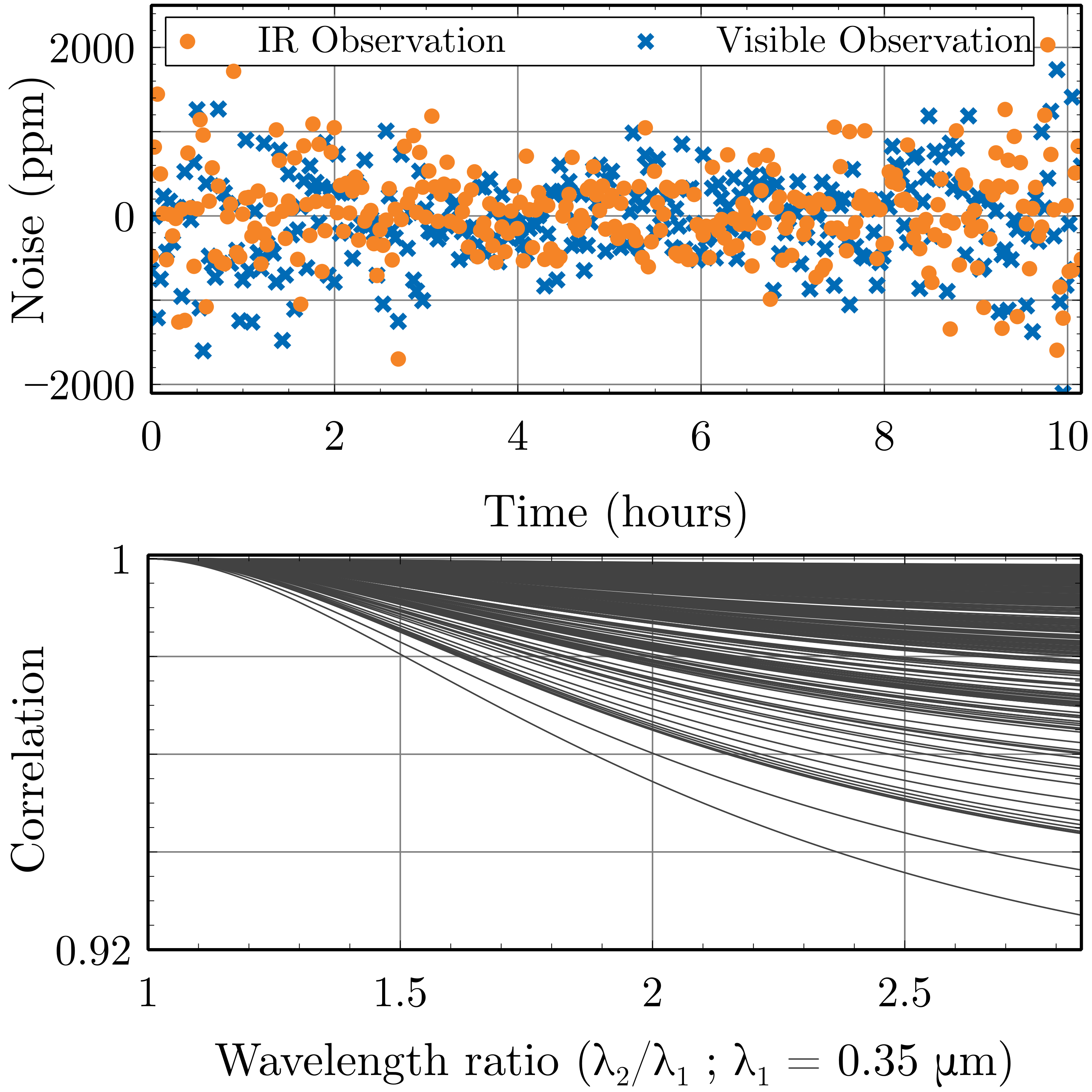}
 %add cn2 profile
 \caption{\textbf{How the correlation between wavelengths changes at each time interval.} Since the correlation coefficient weakly depends on the strength of the vertical turbulence profile, each of these can be well estimated by the correlation at a single altitude, wind speed, and wind direction.}
 \label{fig:correlation_at_timestamp}
\end{figure}

The night of April 29th 2016 from the '2018A' release will be used in this simulation. The data has a temporal resolution of roughly 120s, a vertical resolution of $250$m up to $25$km, and strong variations in the amount of scintillation since the scintillation noise across the total observation period of over 10 hours. A few minor adjustments need to be made to the $\theta(z)$ and $w(z)$ time series for them to be used in the simulation. First, the authors of the dataset communicate that both time-series are typically not sampled uniformly in time. However, when combining the points from separate time-series together, it becomes clear that the general shape of each function with respect to altitude $z$ is stable over time. Therefore this simulation assumes the wind-speed and direction are constant in time. The wind-speed and direction are also not sampled up to the maximum altitude of $25$ km. For the wind-speed, first all measurements of the wind-speed are averaged over the altitude to create one master $v_{\text{avg}}(z)$. Then any points are were not sampled in between the minimum and maximum altitudes are fit with the linear interpolant method in \textsc{numpy}. Finally, the remaining altitudes outsides the min/max are all set to the same value of $5$ m/s. When using this interpolated wind-speed to calculate the scintillation noise, it gives good agreement with the scintillation noise calculated by the authors. An identical procedure is followed for the wind-direction, but the altitudes not sampled outside the minimum and maximum are instead set to the mean of the all the points. This is repeated for both the 'visible' and 'infrared' observation. The resulting scintillation noise and how change in correlation from exposure to exposure is plotted in Figure \ref{fig:correlation_at_timestamp}.

Next, 200 ppm of Poisson noise is added to each light curve. For the chosen exposure time of $120$ seconds and spectral resolution of $200$, this amount of Poisson noise roughly corresponds to a star with a J magnitude of 6. Finally, the Poisson noise, scintillation noise, and transit light curve are summed together to create the final light curve.

To mimic a real transmission spectroscopy reduction, only the radius ratio changes between light curves, while the rest of the transit parameters remain the same. Each light curve is assumed to have no time-correlated noise and so the log-likelihood (for each wavelength index $m$) is
\begin{equation}
\log (\mathcal{L}_{m}) = -\frac{1}{2}\sum_{n=1}^{N}\frac{-1}{2}\left(\frac{y_{n} - T_{mn}}{\sigma}\right)^2 -\frac{1}{2} \log(2\pi \sigma^2)
\end{equation}

\noindent Uniform priors from $0.1$ to $0.15$ for $\RpRs$ are assumed. Then the affine-invariant sampler \textsc{emcee} \citep{Foreman-Mackey2013} is used to sample the posterior and square of the median $\RpRs$ distribution is the recovered radius ratio.

%%%%%%%%%%%%%%%%%%%%%%%%%%%%%%%%%%%%%%%%%%%%%%%%%%

% Don't change these lines
\bsp	% typesetting comment
\label{lastpage}
\end{document}